\documentclass[
    aps,
    pra,
longbibliography,superscriptaddress,amsmath,amssymb,amsfonts, 
    tightenlines,
    twocolumn,
    ]{revtex4-2}

\usepackage[english]{babel}

\usepackage{braket}
\usepackage{booktabs}
\usepackage{array}   
\usepackage{etoolbox}
 \usepackage{times} 
 \usepackage{comment}
 \usepackage[colorlinks=true,linkcolor=teal,citecolor=teal,urlcolor=teal]{hyperref}
 
\usepackage{graphicx}%
\usepackage{float}
\usepackage{amsthm}%
\usepackage[title]{appendix}%
\usepackage{xcolor}%

\newtheorem{theorem}{Theorem}

\begin{document}

\title{Below-threshold error reduction in single photons through photon distillation}

\author{F.~H.~B.~Somhorst}

\affiliation{MESA+ Institute for Nanotechnology, University of Twente, 7500 AE Enschede, The Netherlands}

\author{J.~Saied}

\affiliation{QuAIL, NASA Ames Research Center, Moffett Field, CA 94035, USA}

\author{N.~Kannan}

\affiliation{QuiX Quantum B.V.,  7521 AN Enschede, The Netherlands}

\author{B.~Kassenberg}

\affiliation{QuiX Quantum B.V.,  7521 AN Enschede, The Netherlands}

\author{J.~Marshall}
\affiliation{QuAIL, NASA Ames Research Center, Moffett Field, CA 94035, USA}

\affiliation{USRA Research Institute for Advanced Computer Science, Mountain View, CA 94043, USA}

\author{M.~de Goede}

\affiliation{QuiX Quantum B.V.,  7521 AN Enschede, The Netherlands}

\author{H.~J.~Snijders}

\affiliation{QuiX Quantum B.V.,  7521 AN Enschede, The Netherlands}

\author{P.~Stremoukhov}

\affiliation{QuiX Quantum B.V.,  7521 AN Enschede, The Netherlands}

\author{A.~Lukianenko}

\affiliation{QuiX Quantum B.V.,  7521 AN Enschede, The Netherlands}

\author{P.~Venderbosch}

\affiliation{QuiX Quantum B.V.,  7521 AN Enschede, The Netherlands}

\author{T.~B.~Demille}

\affiliation{QuiX Quantum B.V.,  7521 AN Enschede, The Netherlands}

\author{A.~Roos}

\affiliation{QuiX Quantum B.V.,  7521 AN Enschede, The Netherlands}

\author{N. Walk}

\affiliation{Dahlem Center for Complex Quantum Systems, Freie Universitat Berlin, 14195 Berlin, Germany}

\author{J.~Eisert}

\affiliation{Dahlem Center for Complex Quantum Systems, Freie Universitat Berlin, 14195 Berlin, Germany}

\affiliation{Helmholtz-Zentrum Berlin fur Materialien und Energie, 14109 Berlin, Germany}

\affiliation{Fraunhofer Heinrich Hertz Institute, 10587 Berlin, Germany}

\author{E.~G.~Rieffel}

\affiliation{QuAIL, NASA Ames Research Center, Moffett Field, CA 94035, USA}

\author{D.~H.~Smith}

\affiliation{QuiX Quantum B.V.,  7521 AN Enschede, The Netherlands}

\author{J.~J.~Renema}
\email{j.j.renema@tue.nl}
\altaffiliation{Present address: Department of Applied Physics and Science Education \& Department of Electrical Engineering, Eindhoven University of Technology, P. O. Box 513, 5600 MB Eindhoven, The Netherlands}
\affiliation{MESA+ Institute for Nanotechnology, University of Twente, 7500 AE Enschede, The Netherlands}
\affiliation{QuiX Quantum B.V.,  7521 AN Enschede, The Netherlands}

\maketitle

{\bf Photonic quantum computers use the bosonic statistics of photons to construct, through quantum interference, the large entangled states required for measurement-based quantum computation \cite{briegel2009measurement,FusionBased}. Therefore, any which-way information present in the photons will degrade quantum interference and introduce  errors \cite{forbes2025heralded,rohde2006error,bombin2023increasing}. While quantum error correction can address such errors in principle, it is highly resource-intensive and operates with a low error threshold, requiring numerous high-quality optical components \cite{Roads, QECBasic}.
We experimentally demonstrate scalable, optimal photon distillation as a substantially more resource-efficient strategy to reduce indistinguishability errors in a way that is compatible with  fault-tolerant operation. Photon distillation is an intrinsically bosonic, coherent error-mitigation technique which exploits quantum interference to project single photons into purified internal states, thereby reducing indistinguishability errors at both a higher efficiency and higher threshold than quantum error correction \cite{sparrow2018quantum,marshall2022distillation, somhorst2025photon, saied2025general}. We observe unconditional error reduction (i.e., below-threshold behaviour) consistent with theoretical predictions, even when accounting for noise introduced by the distillation gate, thereby achieving actual net-gain error mitigation under conditions relevant for fault-tolerant quantum computing. We anticipate photon distillation will find uses in large-scale quantum computers. We also expect this work to inspire the search for additional intrinsically bosonic error-reduction strategies, even for fault-tolerant architectures.}

\section{Introduction}\label{sec:intro}
Quantum computing stands to revolutionise fields ranging from cryptography and the study of quantum materials to optimisation and medicine \cite{montanaro2016quantum,GrandChallenge,MindTheGaps}. To realise these promises, quantum noise and loss of coherence remain the main adversaries, and hence low-noise operation of quantum devices is widely seen as the major goal of the field \cite{MindTheGaps}. \emph{Quantum error correction}  (QEC) \cite{QECBasic}, which uses redundancy to protect fragile quantum information, is often regarded as the primary means to achieve this goal. This method carries a steep price, however: the extensive redundancy required by QEC given current error rates introduces substantial resource 
overheads that remain difficult to accommodate in near-term implementations, despite substantial 
recent progress \cite{bluvstein2025fault,google2025quantum,daguerre2025experimental}.  

\textit{Quantum error mitigation} comprises all techniques to reduce errors short of full QEC. In its original formulation, error mitigation takes the form of classical pre- or postprocessing \cite{RevModPhys.95.045005,ZNE,PhysRevA.104.052607}, reducing errors at the cost of exponential sample complexity in the volume of the quantum circuit \cite{ErrorMitigationObstructions,ErrorMitigationObstructionsOld}. More recent approaches have pursued \emph{coherent} error mitigation \cite{PRXQuantum.4.010303,PhysRevX.11.041036,PhysRevX.11.031057,EmilioPaper,VirtualChannelPurification,marshman2018passive,marshman2024unitary}, where quantum operations between multiple copies of the state are allowed, as well as classical postprocessing. Among such protocols, distillation \cite{PhysRevX.11.041036,PhysRevA.86.052329,daguerre2025experimental} focusses on preparing, out of several noisy initial states, a single copy of a target state with reduced noise. These schemes borrow certain elements from QEC \cite{Bennett, Roads}, such as featuring an error threshold in a sense and reducing errors through redundancy and coherent operations, yet they remain significantly easier to realise, as no actual encoding is required. This has made notions of distillation essential in several proposed quantum communication \cite {Kalb} and computing \cite{Pattison} architectures.  

Such techniques are particularly important for photonic quantum computing, since this modality demonstrates its technological advantages at large scale, making efficient resource use an important consideration. The challenges of photonic quantum computing differ sharply from those of other modalities: matter-based systems must be protected from unwanted interactions, while photons are noninteracting particles. Entanglement can nevertheless be built up between such particles by exploiting their exchange statistics, where particle indistinguishability enforces symmetrisation of the wave function \cite{fox2006quantum}. Such quantum interference can then be used in the context of measurement-based quantum computation to generate a cluster state allowing us to reduce all computation to single-particle measurements on that state \cite{briegel2009measurement}. The probabilistic nature of this process means that in photonic quantum computers, state generation accounts for the vast majority of the computing hardware. Errors arise when photons unintentionally carry which-path information, degrading the quantum interference \cite{shchesnovich2015partial}. Such \textit{photon indistinguishability errors} both reduce the probability of successful generation of the resource states \cite{forbes2025heralded} and result in computation errors \cite{rohde2006error,bombin2023increasing}, which has led to substantial efforts to reduce them  \cite{PsiQuantumPlatform,Xanadu}.  

In this work, we experimentally demonstrate the distillation of single photons \cite{sparrow2018quantum}, using an optimal, scalable protocol \cite{marshall2022distillation, somhorst2025photon, saied2025general} directly implementable in a fault-tolerant photonic quantum computer. We utilise quantum interference in an integrated silicon-nitride photonic optical circuit \cite{Taballione2023modeuniversal,degoede} with high precision to entangle multiple copies of a state with finite partial distinguishability, and we make a measurement projecting to a single copy of a distilled state. We find a reduction in photon indistinguishability error by a factor of $\mathcal{R}~=~2.2$, demonstrating that our protocol is running below threshold. Even taking into account the excess noise of our gate, we find an error reduction of $\mathcal{R}~=~1.2$, demonstrating net-gain error mitigation under realistic conditions, showing this protocol is ready to be implemented in quantum computers.

\begin{figure*}[t!]
 \includegraphics[width=16.5cm,   keepaspectratio]{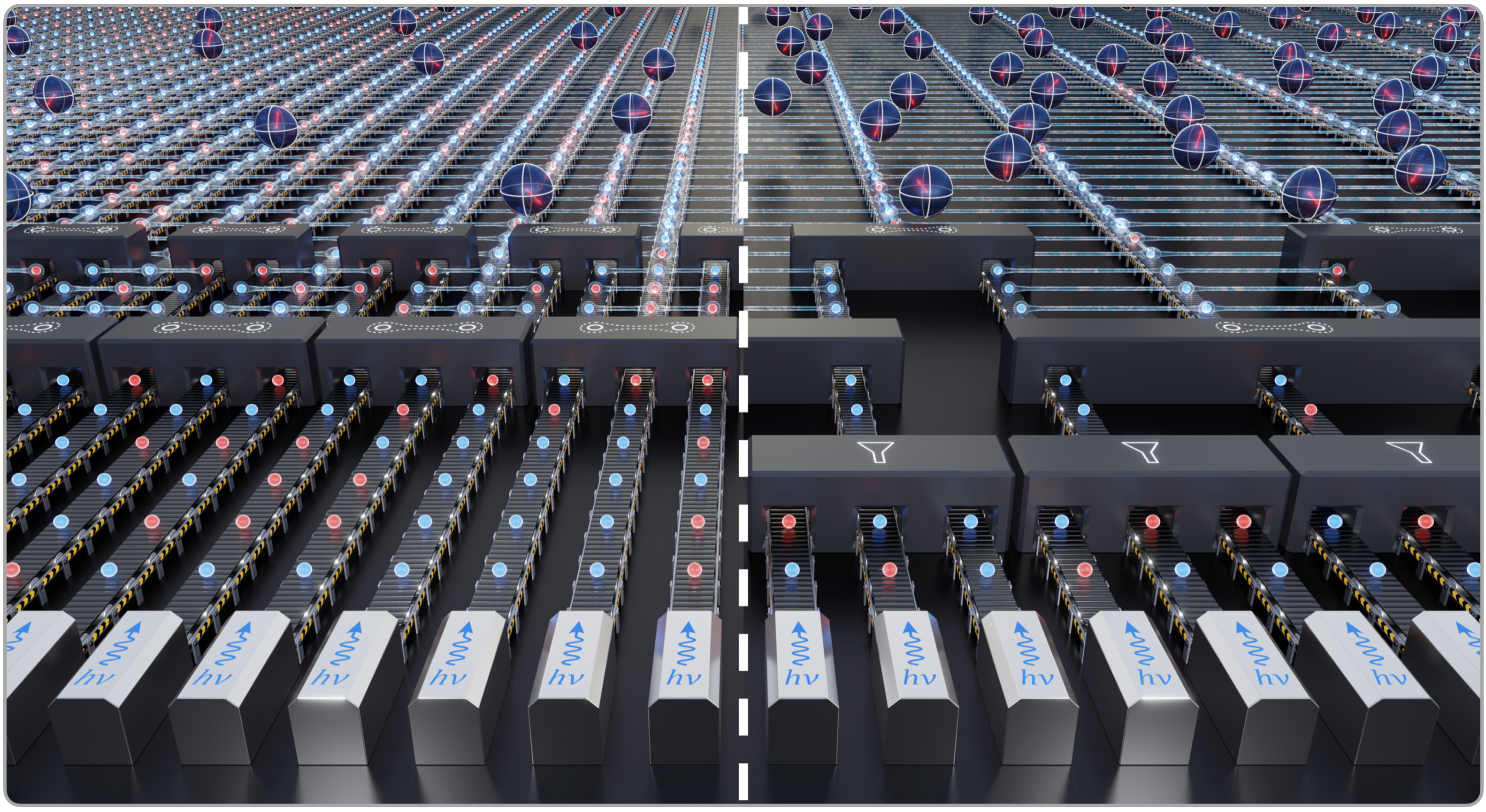}
    \caption{\textbf{Schematic overview of measurement-based quantum computing without (left) and with (right) photon distillation.} In both versions, single photons (depicted as particles) are consolidated into small resource states which are probabilistically combined into a larger resource state, on which a computation is performed by measurement (not shown). Photons carrying indistinguishability errors, shown as red particles, accumulate in the cluster, necessitating error correction to produce logical qubits (shown as Bloch spheres above the cluster). Photon distillation results in a smaller cluster able nonetheless to support more logical qubits due to the suppressed indistinguishability error.}  
    \label{fig:mainr1Artist}
\end{figure*}

We calculate the effects of implementing this gate in photonic quantum computers, considering currently existing sources \cite{paesani2020near, zhai2022quantum, PsiQuantumPlatform}. Optimising the gate for the lowest-indistinguishability photon source reported so far, we estimate a reduction in the total number of photon sources required to construct a logical qubit by a factor of 4. This solution utilises a combination of QEC and photon distillation, emphasizing the complementarity of the two. Since photon sources are by far the most numerous components of the quantum computer, this combination directly corresponds to an equivalent reduction in the bill of materials (Figure\ \ref{fig:mainr1Artist}). Furthermore, we find that with photon distillation, quantum-dot single-photon sources, which suffer strongly from indistinguishability errors between independent sources \cite{zhai2022quantum}, can be reduced to below the error correction threshold.  

\section{Optimal distillation of indistinguishable photons}

Photon distillation has been introduced by Sparrow et al. \cite{sparrow2018quantum}, being a direct consequence of bosonic bunching; that is, the tendency of single bosons in multiple modes to coalesce into a single mode under quantum interference \cite{fox2006quantum}. This protocol captures the intuition of photon distillation, namely that bunching will occur with higher probability if the photons are indistinguishable. Therefore, conditional on bunching, there is a higher posterior probability of the photons being indistinguishable.

Although it has been experimentally implemented in a two-photon version \cite{faurby2024purifying,hoch2025optimal},
this protocol is costly because it either requires concatenation of multiple instances of the protocol or probabilistic coalescing of a large number of photons, both of which are inefficient. To remedy this inefficiency, a three-photon protocol has been found that does not rely on probabilistic coalescing but instead directly projects into the single-photon state \cite{marshall2022distillation}. It was found that this protocol is the smallest version of a family of protocols which uses Fourier interferometers of arbitrary size and can, therefore, operate on arbitrary photon numbers \cite{somhorst2025photon, saied2025general}, removing also the need for concatenation.  
All these protocols have in common that they take $N$ copies of a single-photon state, which in the absence of other experimental noise can be parameterised (see Appendix \ref{app:PhotonErrorModel}):
\begin{equation}
    \rho(\epsilon_{\text{indist}}) = (1-\epsilon_{\text{indist}})|1\rangle \langle 1| + \epsilon_{\text{indist}} |\tilde{1}\rangle \langle \tilde{1}|,
    \label{eq:4_2}
\end{equation}
where $|1\rangle$ is the target state vector of the photon, which is assumed to be common among all $N$ copies, and $|\tilde{1}\rangle$ is a state vector with a different internal mode structure (e.g., polarisation, spectrum), which is to be rejected. Coherent interference  on a linear optical network $\cite{kok2007linear}$ followed by measurement of all but one of the modes probabilistically projects the final mode unto a single photon with reduced indistinguishability error $\epsilon_\text{indist}$. 

The efficiency of these schemes can be measured by the probability of success (i.e., the probability that a detection pattern occurs which heralds a reduced-error photon) and the error reduction (i.e., the degree to which the error is reduced when such a pattern is observed). Although the optimality in the success probability of the scheme of References \cite {saied2025general, marshall2022distillation, somhorst2025photon}, which 
consumes $4N$ photons on average, is still an open problem, we prove (see Appendix \ref{App:Proof}) that the error reduction achieved using Fourier matrices, expressed as
\begin{equation}
	\epsilon_{\text{indist}}^\prime = \frac{1}{N} \epsilon_{\text{indist}} + \mathcal{O} \left( \epsilon_{\text{indist}}^2 \right),
    \label{eq:OptimalFourierAsympt}
\end{equation}
is optimal over all unitaries, where $\epsilon'_{\text{indist}}$ is the indistinguishability error after distillation.

\section{Implementation of a photon-distillation gate}

\begin{figure}[h!]
   \centering
   \includegraphics[width=1.\linewidth]{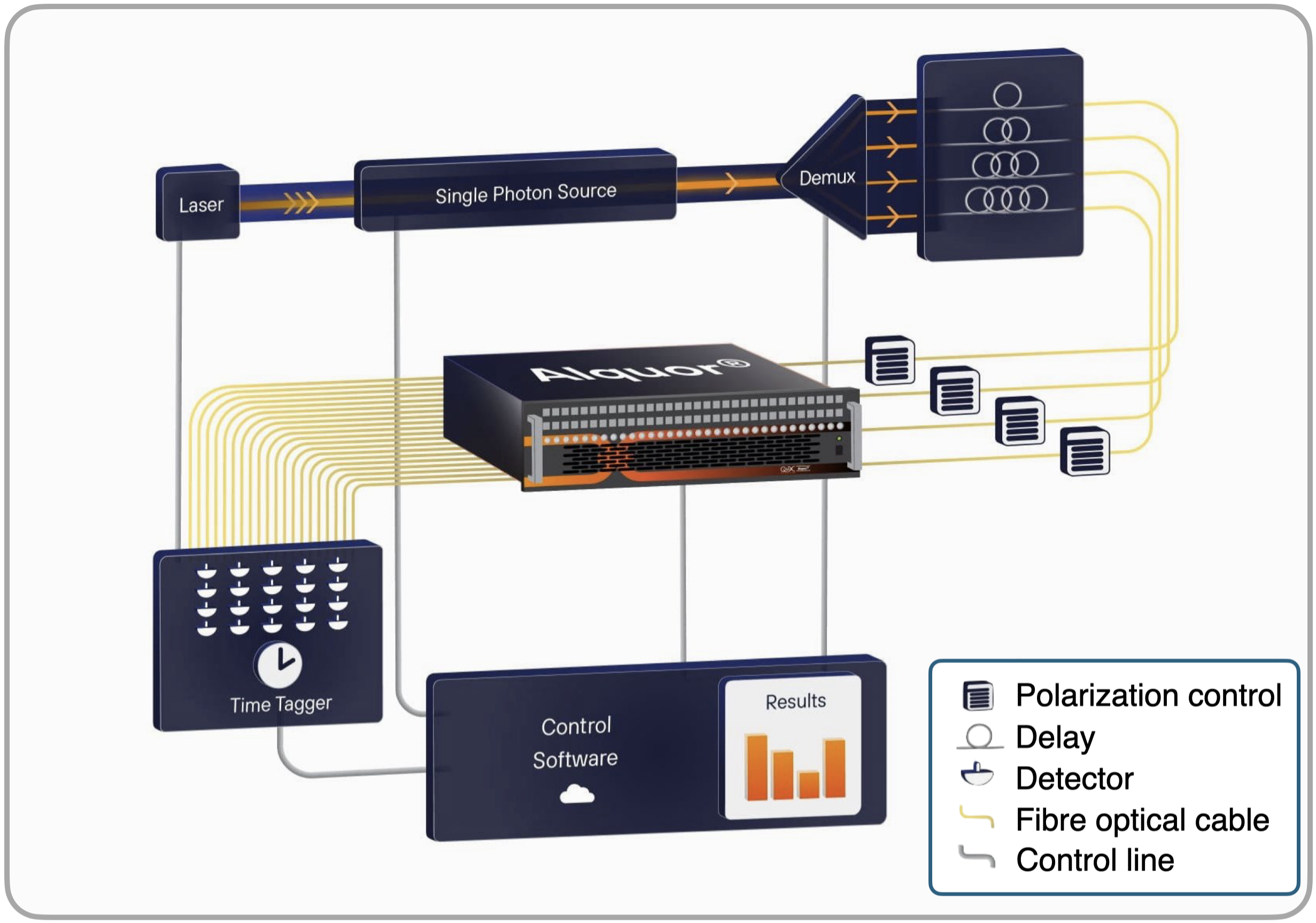}
    \caption{\textbf{Schematic of the experimental platform.} A laser resonantly excites an InGaAs quantum-dot single-photon source, whose photons are demultiplexed from the time domain into spatial modes and synchronised using fibre delays. A 20-mode quantum photonic processor coherently manipulates the photons according to user-specified transformations. Each output mode is individually monitored by a superconducting nanowire single-photon detector, whose signal is read out by a time-tagger to record coincident photon arrivals.}
   \label{fig:bia_functional}
\end{figure}

Our experimental photon-distillation gate consists of a programmable quantum photonic processor \cite{Taballione2023modeuniversal} for coherent manipulation and of superconducting na\-no\-wi\-re single-photon detectors for measurement. The photonic processor, implemented in silicon-nitride integrated photonics \cite{roeloffzen2018low}, consists of 20 modes with on average 3.79 dB insertion loss per mode and programmable pairwise interactions via thermo-optically tuneable Mach–Zehnder unit cells. We programme out the distillation gate and associated calibration measurements in this chip (see Appendices \ref{App:ExpMethods} and \ref{App:DataProcessing}, respectively).

\begin{figure*}[t!]
    \centering
    \includegraphics[width=17cm,  keepaspectratio]{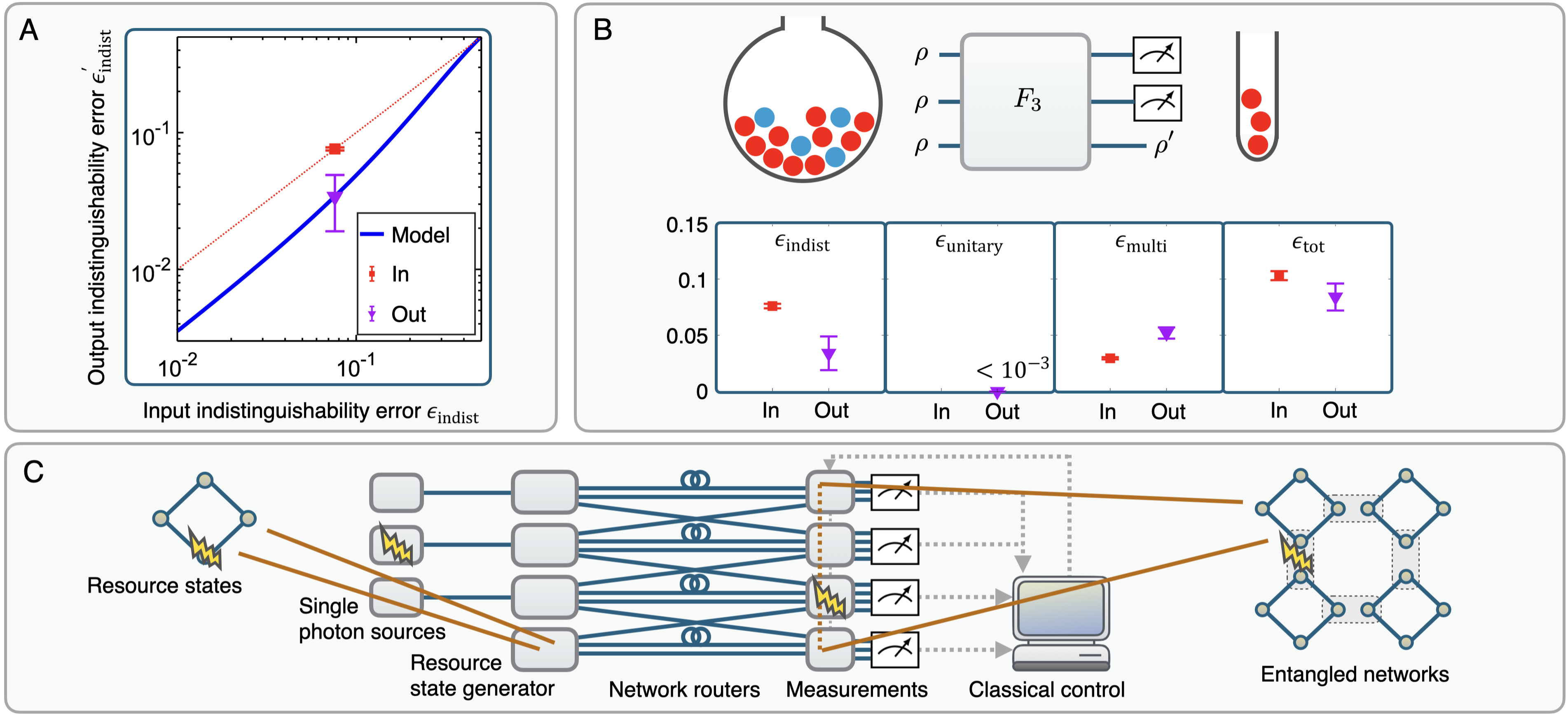}
    \caption{\textbf{Characterisation of the implemented photon-distillation gate.} A) Measured indistinguishability errors and the corresponding improvements predicted by a numerical model based on the characterisation of the implemented linear interferometric transformation. B) Schematic of the implemented photon-distillation gate (top) and the characterised indistinguishability errors $\epsilon_{\text{indist}}$ of the raw input photons $\rho$ and the distilled output photons $\rho^\prime$ (bottom). Gate success is indicated by simultaneous single-photon detection at both measured outputs. Excess noise contributions are expressed as effective indistinguishability errors (interferometric transformation infidelity $\epsilon_{\text{unitary}}$, multiphoton error $\epsilon_{\text{multi}}$), which together induce the total indistinguishability error $\epsilon_{\text{tot}} \simeq \epsilon_{\text{indist}} + (1-\epsilon_{\text{indist}})\epsilon_{\text{multi}}$ \cite{somhorst2025extracting}. C) Simplified depiction of the preparation of a measurement-based photonic quantum computer. The lightning bolts illustrate the different steps where indistinguishability errors  can compromise the computation. Throughout this figure, error bars indicate 95\% confidence intervals.}
    \label{fig:mainr2}
\end{figure*}

To measure the performance of the photon-distillation gate, we prepare photons outside the processor from a resonantly driven semiconductor InGaAs quan\-tum-dot source at 939.5 nm \cite{uppu2021quantum}, which is excited at a repetition frequency of $\tau_r~=~{79.1}$ MHz. These photons are routed to the processor via a demultiplexer, which converts a stream of single photons into batches of four simultaneously arriving photons, each in a separate spatial mode, and which synchronises the time of arrival of the photons. Figure\ \ref{fig:bia_functional} summarises this experimental setup.

We then measure error reduction in photon distillation, shown in Figure\ \ref{fig:mainr2} A. Using the processor's programmability, we implement a set of \textit{in situ} calibration measurements, which allow us to measure the indistinguishability error in the presence of other noise sources \cite{somhorst2025extracting,gonzalez2025two}. We perform a three-photon version of the scalable distillation protocol, using a fourth photon as a reference to measure the distinguishability via Hong-Ou-Mandel interference \cite{hong1987measurement}. We measure the indistinguishability error both before and after distillation, finding errors of $\epsilon_{\text{indist}}~=~0.076 $ (95\% CI: 0.074 – 0.078) and $\epsilon^\prime_{\text{indist}}~=~0.034$ (95\% CI: 0.019 – 0.049), corresponding to a 2.2-fold reduction. This improvement matches the predicted reduction from Eq.\ (\ref{eq:OptimalFourierAsympt}), considering finite-size effects in the error \cite{marshall2022distillation}, demonstrating the protocol and confirming that this protocol is operating below threshold at a level unachievable by earlier experiments \cite{faurby2024purifying,hoch2025optimal}. 

However, this observed reduction in indistinguishability error does not, by itself, make the distillation circuit useful for quantum computers. Analogous to the notion of fault-tolerant operation in QEC, we must establish \textit{net-gain} error mitigation, ensuring that the achieved error reduction is not outweighed by the excess noise it introduces. Therefore, a full characterisation of the excess noise sources is necessary. To facilitate a like-for-like comparison, we express all of these noise sources in terms of effective indistinguishability errors (i.e., how much higher the indistinguishability error would appear to be in light of this excess noise). 

Figure \ref{fig:mainr2} B shows the result of these characterisations. We first consider the unitary error $\epsilon_{\text{unitary}}$, representing the degree of accuracy with which we control our chip. We determine the interferometric transformation fidelity $\mathcal{F} = \frac{1}{3} {\rm tr}   (U_{\text{th}}^\dagger U_{\text{exp}} ) =
99.8\%$, where $U_\text{th}$ and $U_\text{exp}$ are the set and get values for the optical transformation implemented by the gate (see Appendix \ref{app:matrixCharacter}). We propagate this transformation infidelity into an effective indistinguishability error of $7 \times 10^{-4}$, which is negligible compared to all other errors.\footnote{The effective indistinguishability error is quantified as the excess output error relative to an ideal Fourier transform, with imperfect implementations modelled using the simulation techniques of Reference \cite{somhorst2025photon}.} 

Next, we characterise to multiphoton noise. Quantum dots occasionally produce multiple photons, for example by scattering a pump photon or through re-excitation. Since these photons differ in time or frequency from the intended mode, they are best modelled as fully distinguishable \cite{ollivier2021hong}. In the case of multiphoton emission combined with optical loss, these photons take the place of a photon occupying the intended mode, thereby elevating the indistinguishability error \cite{somhorst2025extracting}. We measure the effect of such substitutions as $\epsilon_{\text{multi}}~=~0.030$ (95\% CI: 0.029  – 0.031) before the gate, which increases to $\epsilon^\prime_{\text{multi}}~=~0.052$ (95\% CI: 0.047  – 0.057) after the gate. Since our method for arriving at the indistinguishability reduction considers only these sources of error and since our observed error reduction agrees fully with theory, we conclude that there are no other significant sources of noise in the experiment.

Putting all effects together, we find $\epsilon_{\text{tot}}~=~0.103$ (95\% CI: 0.099  – 0.107) for the raw input photons and $\epsilon_{\text{tot}}^\prime~=~0.084$ (95\% CI: 0.072  – 0.096) for the distilled output photons. This 1.2-fold reduction demonstrates that the photon-distillation gate achieves net-gain error mitigation. 

We briefly discuss the impact of these noise sources on the applicability of the gate in photonic quantum computers. Our results on unitary noise demonstrate that chip control is not the limiting factor on distillation. While the dependence on chip control does increase with photon number, extrapolation from current error rates indicates that this error will remain negligible until $N \gtrsim 40$.\footnote{The three-photon transformation infidelity corresponds to an effective indistinguishability error $\epsilon_{\text{unitary}}~=~7~\times~10^{-4}$, implying an effective single-photon fidelity of $(1-\epsilon_{\text{unitary}})^{\frac{1}{3}}$ under independent error accumulation. Extrapolation to an $N$-photon process gives $\epsilon_{\text{unitary}}^\prime = 1 - (1-\epsilon_{\text{unitary}})^{\frac{N}{3}}$, which for $N~\gtrsim~40$ is comparable in magnitude to other effective indistinguishability errors ($\epsilon_{\text{unitary}}^\prime~\sim~10^{-2}$).} Multiphoton noise can be reduced to the $10^{-6}$ level by introducing suitable engineering of the excitation conditions  \cite{nelson2025noise}, which is far below the threshold of state-of-the-art quantum error correcting codes. For heralded single-photon sources, a heralding detector resolving photon numbers can reduce multiphoton errors to an arbitrary degree \cite{sempere2022reducing}. In what follows, we will therefore assume that this excess noise has been reduced to negligible level.

\section{Resource reduction in quantum computers}

We assess the impact of photon distillation on resource use in photonic quantum computers with a simplified model of resource use  \cite{somhorst2025photon}, computing the number of single photons required to construct a logical qubit using the surface code (see Appendices \ref{app:PhotonResourceEstimate} and \ref{app:ScalingOverview}). To maintain scalability, we require that the size of the distillation circuit is constant in the size of the computation. Since photon sources are the first component in a photonic quantum computer, their number is a good proxy for the resource cost of the computer as a whole. Indistinguishability errors manifest themselves as failed steps in the computation throughout the process of resource state generation and measurement (see Figure\ \ref{fig:mainr2} C). We make the assumption that photon indistinguishability affects only the computation at readout. In such a model, indistinguishability errors directly translate into Pauli errors \cite{rohde2006error}, which must be corrected by QEC. This model underestimates the impact of photon distillation, because distillation occurs before resource state formation and QEC occurs after, and the effects of indistinguishability on resource state generation are known to be substantial \cite{forbes2025heralded}. The trade-off here is between consuming photons in the distillation protocol and reducing the required number of physical qubits to create a logical qubit through the error reduction.  

\begin{figure}[H]
    \centering
    \includegraphics[width=1\linewidth]{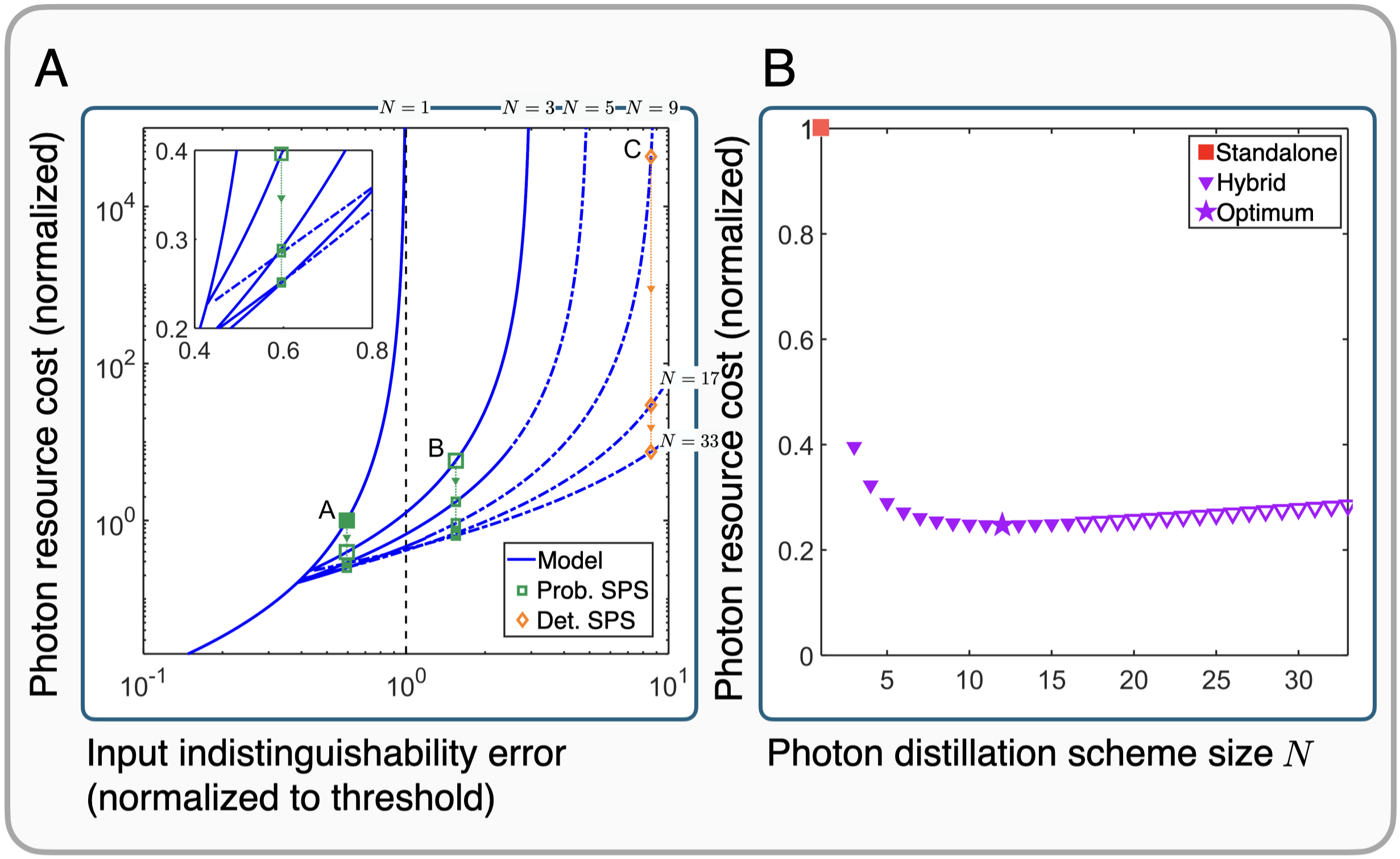}
    \caption{\textbf{Effect of distillation on quantum computer resource cost.} A) Resource use as a function of indistinguishability error (normalised to threshold) for different sizes $N$ of the photon-distillation scheme. Each distillation isoline shows the number of photons required to achieve a logical error $p_{\text{L}} = 10^{-10}$, with $N = 1$ representing no distillation, normalised to the requirements of the currently best-performing source. Solid lines indicate where linear error reduction, as in Eq.\ (\ref{eq:OptimalFourierAsympt}), is valid within 2\%; dashed lines are shown otherwise. The points mark three state-of-the-art photon sources (SPS) [A] \cite{PsiQuantumPlatform}, [B] \cite{paesani2020near}, and [C] \cite{zhai2022quantum}, with green squares marking probabilistic sources and orange diamonds marking deterministic sources. Inset: zoom-in to the area around source [A]. B) Resource cost for photon source [A] as a function of photon-distillation scheme size $N$, illustrating that a complementary $N = 12$ photon-distillation scheme can reduce costs by a factor of four, relative to standalone QEC. Filled markers indicate where linear error reduction is valid within 2\%; open markers are shown otherwise.}
    \label{fig:ResourceCostSize}
\end{figure}

Figure\ \ref{fig:ResourceCostSize} A shows resource use as a function of physical error probability for different protocol sizes $N$. To generate each line, we fix the number of photons involved in the distillation step $N$ and compute the number of physical qubits at a given physical error $p_{\text{error}}$ required to achieve a constant logical error of $p_{\text{L}}~=~10^{-10}$, where $N~=~1$ represents no distillation (i.e., pure error correction). The solid part of each line indicates where linear error reduction as given by Eq.\ (\ref{eq:OptimalFourierAsympt}) is valid to the $2\%$ level, the dashed lines indicate where correction terms play a larger role. The three points indicate three state-of-the-art photon sources, with colour coding for source type as either probabilistic based on nonlinear optics or as a deterministic solid-state source.  

There are three regimes: a regime from $p_{\text{error}} = 0$ to $p_{\text{error}} = 0.39 p_{\text{L}}$ where pure error correction is optimal; a regime above that but below the error correction threshold, where fault-tolerant operation is possible without distillation, but where there is a resource reduction; and a regime above the error correction threshold but below the photon-distillation threshold, where distillation is necessary to achieve fault-tolerant operation. The source from [A], which represents the lowest-indistinguishability error probabilistic source, falls in the error-reduction regime, and the deterministic solid-state source from [C] falls in the regime where fault-tolerant operation is possible only with distillation. 

Figure \ref{fig:ResourceCostSize} B shows the resource reduction possible using photon distillation as applied to the lowest-indistinguishability error source currently reported. We find that a maximum resource reduction to a factor of 0.25 is achieved at $N~=~12$, a size achievable with current technological capabilities.

\section{Discussion}
In this work, we have presented experimental results on
a protocol of photon distillation that leads to a 
net-gain error reduction. We emphasise the extent to which our model of resource use in a quantum computer is skewed to underestimate photon distillation, since it considers only errors which occur at the end of the computation. A full study of error propagation in photonic quantum computers including correlated errors and out-of-code errors is necessary to more accurately assess the effect of distillation. It is likely that the regime where distillation is useful will expand further below threshold and that the resource reduction will increase. 

Second, we address the main source of out-of-code errors, namely photon loss. We note that, since the size of the distillation circuit does not depend on the number of qubits, distillation imposes an additional constant optical loss on the circuit. We can assess the effect of photon loss on the feasibility of photon distillation as a protocol by observing that the distillation gate is very similar to the other gates required for measurement-based quantum computing (MBQC), consisting as it does of a linear circuit, some heralding detectors, and a fast switch which routes the resulting state based on the measured heralding pattern. Hence, we can assume it can be implemented with the same optical loss as the other gates, such as state generation and fusion. In conventional MBQC, each photon passes through roughly $G~=~6-12$ such gates, depending on the precise protocol. Neglecting higher-order terms, the tolerable loss per component $l$ is reduced to $l^\prime~=~\frac{G}{G+1}l$, where $l^\prime$ is the tolerable loss per component after distillation; for reasonable values of $G$, we expect a $<15\%$ reduction in component-level loss error budget, resulting in an $N$ times higher overall budget for indistinguishability error. Since the interchange between indistinguishability errors and loss errors is roughly linear, this increased budget opens up a substantial fraction of the parameter space for fault-tolerant operation.

Finally, our work raises the question of the extent to which other fully bosonic techniques can be used to improve the efficiency of a photonic quantum computer, for example by finding intrinsically bosonic distillation schemes for other errors, such as multiphoton errors. This result also emphasises the degree to which techniques from the noisy intermediate-scale quantum (NISQ) and fault-tolerant regimes flow into each other, since indistinguishability errors were first fully studied in the context of NISQ devices \cite{shchesnovich2015partial}. It also further suggests exploring the \emph{middle ground} between quantum error mitigation and correction.

\subsection*{Acknowledgements}
F.~H.~B.~S. is supported by the Photonic Integrated Technology Center (PITC). J.~J.~R. is supported by the project “At the Quantum Edge" of the research programme VIDI, which is financed by the Dutch Research Council (NWO). QuiX is supported by the Purple Nectar 2025 Quantum Challenges project QSHOR of the Materiel and IT command of the Ministry of Defense of the Netherlands, a contribution from the INLITE project of the National Growth Fund program NXTGEN HIGHTECH, and by the European Union within the framework of the European Innovation Council’s Pathfinder program, under the project QuGANTIC, and the H2020-FETOPEN Grant PHOQUSING (GA no.\ 899544).
J.~S. and E.~G.~R. are grateful for support from  DARPA under IAA 8839, Annex 130, and from  NASA Ames Research Center. J.~M.~is thankful for support from NASA Academic Mission Services, Contract No.\ NNA16BD14C. J.~E.~has been supported by the BMFTR (PhoQuant, QPIC-1, PasQuops, QSolid), Berlin Quantum, the Munich Quantum Valley, the Clusters of Excellence ML4Q and MATH+, the DFG (CRC 183 and SPP 2541),
and the European Research Council
(ERC AdG DebuQC).  We thank S.~N.~van den Hoven, P.~M.~Ledingham, H.~Offerhaus, D.~DiVinchenzo and B.~Terhal for discussions. 
The United States Government retains, and by accepting the article for publication, the publisher acknowledges that the United States Government retains, a nonexclusive, paid-up, irrevocable, worldwide license to publish or reproduce the published form of this work, or allow others to do so, for United States Government purposes.

\subsection*{Author contributions}
F.~H.~B.~S., J.~S., J.~M., and J.~J.~R. conceived the experiment. F.~H.~B.~S., J.~S., and J.~M. conducted preliminary simulations. J.~S. constructed the optimality proof. B.~K., M.~d.~G., H.~J.~S, A.~L. , and P.~S. constructed the experimental setup. P.~V., T.~B.~D., A.~R., and D.~H.~S supervised the development of the experimental apparatus. F.~H.~B.~S., N.~K., and J.~S. developed the experimental methodology. N.~K., and B.~K. conducted experiments. F.~H.~B.~S. conducted data analysis. F.~H.~B.~S., J.~S., J.~E., E.~G.~R., and J.~J.~R. interpreted the results. J.~S., and N.~K. validated the data analysis. F.~H.~B.~S., and N.~K. performed data curation. F.~H.~B.~S., N.~W., and J.~J.~R. developed the QEC model. F.~H.~B.~S., N.~K., J.~E., and J.~J.~R. wrote the manuscript. E.~G.~R., and J.~J.~R. provided overall supervision of the research project.

\subsection*{Competing interest declaration}

J.~J.~R.~is a shareholder in QuiX Quantum B.~V. The remaining authors declare no competing interests.

\subsection*{Data availability}
All experimental and simulated data used in this study are available in the 4TU.ResearchData repository (DOI: 10.4121/2ddcf5bd-2c3b-4986-8bba-aef7be5ae9aa). 

\begin{appendices}

\section{Photon indistinguishability error model}\label{app:PhotonErrorModel}
We first introduce the photon indistinguishability error $\epsilon_{\text{indist}}$, which serves as a key performance metric in this work. To describe these stochastic errors in the internal degrees of freedom of the injected photons, we adopt the \textit{orthogonal bad-bit} (OBB) model \cite{sparrow2018quantum}. This model represents a worst-case scenario for multiphoton interference, as the orthogonal (or \textit{bad}) 
components are assumed to be fully distinguishable from all other photons and therefore cannot interfere with them. The OBB model is widely used, for example, in analyses of photonic quantum advantage experiments \cite{renema2018efficient, wang2019boson, renema2021sample}.

Formally, we represent the state of a noisy photon as 
\begin{equation}
    \rho(\epsilon_{\text{indist}}) = (1-\epsilon_{\text{indist}}) \ket{\psi}\bra{\psi} + \epsilon_{\text{indist}} \rho_\perp, 
    \label{eq:OBBMSPS}
\end{equation}
where $\ket{\psi}$ represents the ideal (indistinguishable) single-photon state vector such that $\bra{\psi}\rho\ket{\psi} = 1- \epsilon_{\text{indist}}$. $\rho_\perp$ represents the orthogonal (distinguishable) component, defined such that the single-photon trace purity is given by
\begin{equation}
    {\rm tr}(\rho(\epsilon_{\text{indist}})^2) = (1-\epsilon_{\text{indist}})^2. 
\end{equation}
Experimentally, this purity is equivalent to the Hong–Ou–Mandel visibility measured between two noisy photons \cite{hong1987measurement}. However, under typical experimental conditions, there are additionally loss and multiphoton errors that i) affect the measured visibility and ii) contribute to the photon indistinguishability error. We follow the method outlined in 
Reference\ \cite{somhorst2025extracting} to infer the total error $\epsilon_{\text{tot}}$, which we then use to extract the indistinguishability error by correcting for the measured multiphoton error $\epsilon_{\text{multi}}$,  
\begin{equation}
    \epsilon_{\text{indist}} = \frac{\epsilon_{\text{tot}} - \epsilon_{\text{multi}}}{1 - \epsilon_{\text{multi}}}.
    \label{eq:eps_int_def}
\end{equation}
Here, the multiphoton noise is assumed to be fully distinguishable from the target state vector $\ket{\psi}$, as is appropriate for the single-photon source used in our experiments \cite{ollivier2021hong}.

Second, we consider possible alternative interpretations of our experimental data to assess the robustness of our conclusions. Alternative models for indistinguishability errors often focus on relaxing the assumption that bad photons occupy distinct subspaces within $\rho_\perp$. Other plausible models, such as the random source model \cite{sparrow2018quantum}, allow overlap in occupied error states, leading to additional interference effects. At the extreme limit, one may consider that each noisy photon shares an identical error state, representing the exact opposite of the OBB model: the \textit{similar bad-bit} (SBB) model \cite{saied2024advancing}. In this case, the noisy single-photon state is expressed as
\begin{equation}
    \sigma(\epsilon_{\text{indist}}) = (1-\epsilon_{\text{indist}})\ket{\psi}\bra{\psi} + \epsilon_{\text{indist}} \ket{\phi}\bra{\phi}, 
\end{equation}
where $\ket{\psi}$ is the ideal state vector and $\ket{\phi}$ represents a common error state vector ($\langle \psi | \phi \rangle = 0$). In case of SBB photons, the single-photon trace purity becomes
\begin{equation}
    {\rm tr}(\sigma(\epsilon_{\text{indist}})^2) = (1-\epsilon_{\text{indist}})^2 + \epsilon_{\text{indist}}^2. 
\end{equation}
Table \ref{tab:eps_model} summarises the experimentally 
extracted photon indistinguishability input errors $\epsilon_{\text{indist}}$ and output errors $\epsilon^\prime_{\text{indist}}$ obtained under both the OBB and SBB models. The table also includes the theoretically expected errors \cite{saied2025general}.  Both models yield quantitatively accurate predictions, consistent with the expectation that differences between noisy-photon models become increasingly negligible for high-quality single-photon sources \cite{faurby2024purifying}. This agreement further validates the use of the OBB model and strengthens the conclusions reported in the main text.

\begin{table}[h]
\centering
\caption{Extracted photon indistinguishability error obtained under two interpretative models: The OBB model and the SBB model. Exp. = Experiment, Th. = Theory.} \label{tab:eps_model}
\begin{tabular}{l l l}
\toprule
Model & OBB & SBB \\ 
\midrule
$\epsilon_{\text{indist}}$ & $0.0759$ & $0.0793$  \\ 
$\epsilon_{\text{indist}}^\prime$ (Exp) & $0.0337$ & $0.0329$ \\ 
$\epsilon_{\text{indist}}^\prime$ (Th) & $0.0335$ & $0.0313$ \\ 
\botrule
\end{tabular}
\end{table}

We also consider the effect of assuming uniform photon indistinguishability errors. In our demultiplexing setup, the pulsed photon stream from a quantum dot is split into $N$ synchronised streams, increasing the maximum emission-time separation between photons with larger $N$. Photons emitted with larger temporal separations are generally less identical due to undesired processes such as charge noise and polarisation drift \cite{tomm2021bright}. Consequently, photons used in the reference experiments are likely more indistinguishable than those used in the distillation experiments, implying $\epsilon_{\text{indist}} \leq \epsilon_{\text{indist}}^{(i)}$ for the $i = 1, \dots, N$-th input photon. Thus, our reported input indistinguishability error, 
\begin{equation}
\epsilon_{\text{indist}} \leq \langle \epsilon_{\text{indist}} \rangle = \frac{1}{N}\sum_{i = 1}^{N} \epsilon_{\text{indist}}^{(i)}
\end{equation}
is likely underestimated in the context of the distillation experiment. Therefore, the reported output indistinguishability error, which is approximately proportional to the input error to first order \cite{somhorst2025photon}, is accordingly overestimated with respect to the reported input error: $ \epsilon_{\text{indist}}^\prime = \frac{1}{N} \langle \epsilon_{\text{indist}} \rangle \geq \frac{1}{N} \epsilon_{\text{indist}}$. Since in our analysis assuming uniform photon indistinguishability errors tends to yield underestimation of the genuine indistinguishability improvement, observing a reduction under this assumption strengthens our conclusions.

\begin{figure}[h!]
    \centering
     \includegraphics[width=.95\columnwidth,  keepaspectratio]{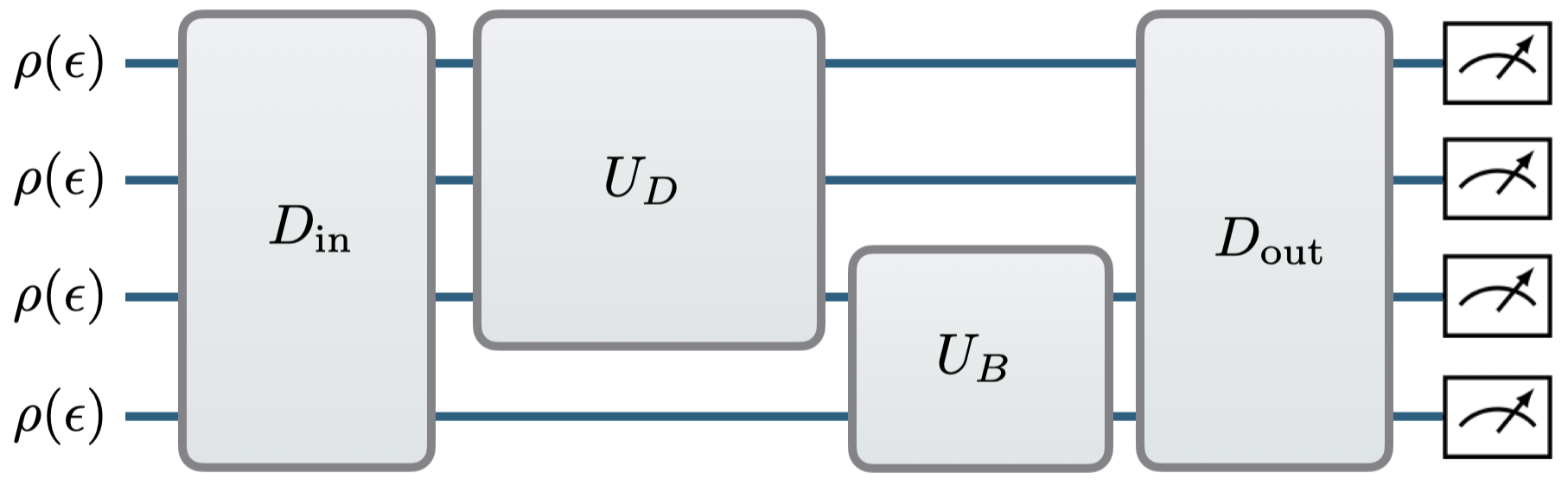}
    \caption{\textbf{Transformation model for non-uniform loss simulation.} $\rho(\epsilon)$ = injected photon with indistinguishability error $\epsilon$; $D_{\text{in (out)}}$ = sub-unitary diagonal matrices capturing non-uniform losses as reported in 
    Eq.\  (\ref{eq:MatrixCharacterFull}); $U_D$ = distillation transformation matrix as reported in Eq.\  (\ref{eq:MatrixU_D_Exp}) $U_B$ = two-mode balanced beam splitter for HOM interference experiment.}
    \label{fig:NonUniformLossSimulation}
\end{figure}

Nonuniform photon losses are known to influence multiphoton interference \cite{clements2016optimal}. From our matrix characterisation measurements, it is evident that some degree of loss imbalance is present in our system, as per Eq.\  (\ref{eq:MatrixCharacterFull}). To assess how these nonuniform losses affect the measured indistinguishability error reduction, we use the transformation model shown in Figure\ \ref{fig:NonUniformLossSimulation} to simulate their impact. Specifically, we employ the experimentally characterised matrix elements of $D_{\text{in}}$ , $D_{\text{out}}$, and $U_D$, as in 
Eqs.\ (\ref{eq:MatrixCharacterFull}) and \ref{eq:MatrixU_D_Exp}, together with a balanced beam splitter $U_B$. Using an experimentally determined input error of $\epsilon_{\text{indist}}~=~0.076$, we compute the expected output indistinguishability error as $\epsilon^\prime_{\text{indist}}~=~0.032$. We find that the predicted output error is $<5\%$ lower than the value expected under uniform loss. Thus, the output error reported in the main text is likely slightly underestimated due to the presence of non-uniform losses. Nevertheless, this small discrepancy is not expected to alter the conclusion that a genuine reduction has been observed.

We assume that the reference and distilled photon states have similar average photon numbers. Under this assumption the measured HOM correlator $g_{\text{HOM}}$  is corrected for multiphoton contributions by using an average multiphoton error, $\bar{\epsilon}_{\text{multi}} = \frac{1}{2} \epsilon_{\text{multi}} + \frac{1}{2} \epsilon_{\text{multi}}^\prime$, to infer the output indistinguishability error \cite{somhorst2025extracting,tsujimoto2021ultra,trivedi2020generation}. This equal-intensity assumption may be violated in practice, for instance because detectors resolving photon numbers are not used in the heralding output ports of the distillation gate. To quantify the effect of a possible intensity imbalance, we follow the method described in Reference\  \cite{tsujimoto2021ultra} and estimate a lower bound on the weighted average multiphoton error given by
\begin{equation}
    \bar{\epsilon}_{\text{multi}} = \frac{1}{2} \zeta \epsilon_{\text{multi}} + \frac{1}{2} \zeta^{-1} \epsilon_{\text{multi}}^\prime,
    \label{eq:g_average_zeta}
\end{equation}
where $\zeta$ is the ratio of average photon numbers. From our measurements we obtained $\epsilon_{\text{multi}}~=~0.030$ and $\epsilon_{\text{multi}}^\prime~=~0.052$. Numerical analysis shows that $\bar{\epsilon}_{\text{multi}}$ is overestimated for $1 < zeta < 1.76$. Overestimating $\bar{\epsilon}_{\text{multi}}$  leads to an underestimation of the output indistinguishability error, which could result in a false indication of successful indistinguishability error reduction. Underestimating 
$\bar{\epsilon}_{\text{multi}}$ only strengthens our conclusion. Even in the worst-case scenario ($\zeta~\approx~1.33$) we find a genuine $\bar{\epsilon}_{\text{multi}}~=~0.0391$, which is only 4\% smaller than the estimation $\bar{\epsilon}_{\text{multi}}~=~0.0408$ used in the main analysis. We conclude, therefore, that the reported indistinguishability error reduction is robust against average photon-number imbalance.

\section{Derivation of optimal asymptotic indistinguishability error suppression}\label{App:Proof}
In what follows, we prove that the Fourier transform achieves optimal asymptotic suppression of the photon indistinguishability error. Specifically, we prove that for any $N \times N$ unitary transformation $U$ implemented by a linear-optics multiport interferometer, the distilled error $\epsilon_{\text{indist}}^\prime$ cannot be reduced below $\epsilon_{\text{indist}} / N$ under the noisy input state $\rho(\epsilon_{\text{indist}})^{\otimes N}$. For brevity, we write $\epsilon_{\text{indist}}$ as $\epsilon$. 

\begin{theorem}[Optimality of the Fourier transform]
For any $N \times N$ linear-optical unitary mode transformation matrix $U$, the distilled error is lower bounded by
    \begin{equation}
         \epsilon^\prime  \geq \frac{\epsilon}{N} + O(\epsilon^2).
    \end{equation}
   \label{th:ErrorSupressionBound}
\end{theorem}
In particular, this bound implies that the Fourier transform is asymptotically optimal for photon distillation up to subleading corrections.\footnote{However, its applicability does not imply that the Fourier transform is the only optimal transformation, as the Hadamard transformation also saturates the same asymptotic bound  \cite{saied2025general, somhorst2025photon}.} 

\begin{proof}
In the asymptotic low-error regime, the product state of noisy single-photon inputs can be decomposed as follows \cite{moylett2019classically}:
\begin{equation}
    \begin{split}
	\rho(\epsilon)^{\otimes N} & = (1 - N \epsilon) {\ket{\psi}\bra{\psi}}^{\otimes N} \\ &+  \epsilon \sum_{k=1}^{N}{\ket{\psi}\bra{\psi}}^{\otimes k-1}\otimes{\rho_\perp}\otimes{\ket{\psi}\bra{\psi}}^{\otimes N-k} + \mathcal{O}\left( \epsilon^2 \right),
    \end{split}
\end{equation}
or equivalently as (higher-order terms omitted)
\begin{equation}
    \begin{split}
	\rho(\epsilon)^{\otimes N} = (1 - N \epsilon) {\ket{\phi_0}\bra{\phi_0}} +  N \epsilon \left( \frac{1}{N} \sum_{k=1}^{N}{\ket{\phi_k}\bra{\phi_k}} \right),
    \end{split}
\end{equation}
where ${\ket{\phi_k}\bra{\phi_k}} = {\ket{\psi}\bra{\psi}}^{\otimes k-1}\otimes{\rho_\perp}\otimes{\ket{\psi}\bra{\psi}}^{\otimes N-k}$; that is, the photon injected into multi-port mode $k$ is fully distinguishable, while all other photons remain fully indistinguishable.

The total probability of producing a heralding pattern under transformation $U$ can be expressed as
\begin{equation}
    \begin{split}
    P_U(\text{herald}) &= P_U\left( \text{herald}\mid{\textstyle \ket{\phi_0}\bra{\phi_0}} \right)P_U\left( { \textstyle \ket{\phi_0}\bra{\phi_0}} \right) \\ &+ P_U \left( \text{herald}\mid{\scriptstyle \frac{1}{N} \sum_{k=1}^{N}{\ket{\phi_k}\bra{\phi_k}} } \right) P_U\left( {\scriptstyle \frac{1}{N} \sum_{k=1}^{N}{\ket{\phi_k}\bra{\phi_k}} }  \right) \\ &
    = p_0 + \mathcal{O}\left( \epsilon \right),
    \end{split}
\end{equation}
where $p_0 = P_U\left( \text{herald}|{\textstyle \ket{\phi_0}\bra{\phi_0}} \right)$ represents the heralding probability for fully indistinguishable photons.
Hence, the new distilled error $\epsilon^\prime$ can be expressed as
\begin{equation}
    \begin{split}
        \epsilon^\prime & = P_U \left( \text{error} | \text{herald} \right) \\ &
        = \frac{P_U \left( \text{error} \cap \text{herald} \right)}{P_U \left( \text{herald} \right)} \\&
        = \frac{P_U \left( \text{error} \cap \text{herald} | {\scriptstyle \frac{1}{N} \sum_{k=1}^{N}{\ket{\phi_k}\bra{\phi_k}} } \right) P_U \left( {\scriptstyle \frac{1}{N} \sum_{k=1}^{N}{\ket{\phi_k}\bra{\phi_k}} } \right) }{P_U \left( \text{herald} \right)} \\&
        = \frac{N P_U \left( \text{error} \cap \text{herald} | {\scriptstyle \frac{1}{N} \sum_{k=1}^{N}{\ket{\phi_k}\bra{\phi_k}} } \right)}{p_0}\epsilon + \mathcal{O} \left( \epsilon^2 \right).
    \end{split}
\end{equation}
We proceed to derive a lower bound on $P_U(\cdot)$. First, we consider the decomposition
\begin{equation}
	\ket{\phi_k} = \frac{1}{\sqrt{N}}\ket{\phi_+} + \ket{\phi_{k,-}},
\end{equation}
where 
\begin{equation}
	\ket{\phi_+} = \frac{1}{\sqrt{N}} \sum_{j=1}^{N} \ket{\phi_j}
    \label{eq:pseudoindiststate}
\end{equation}
is independent of $k$, and the latter component,
\begin{equation}
	\begin{split}
	\ket{\phi_{k,-}} & = \ket{\phi_k} - \frac{1}{\sqrt{N}}\ket{\phi_+}  
		= \ket{\phi_k} - \frac{1}{N} \sum_{j = 1}^{N} \ket{\phi_j}
	\end{split}
\end{equation}
is unnormalised. Importantly, this decomposition satisfies
\begin{equation}
	\sum_{k=1}^{N} \ket{\phi_{k,-}} = 0.
\end{equation}
With this decomposition, the single-error mixed state can be expressed as
\begin{equation}
    \begin{split}
        \frac{1}{N} &\sum_{k=1}^{N} {\ket{\phi_k}\bra{\phi_k}}  = \\
        &\frac{1}{N} \sum_{k=1}^{N} \left( \frac{1}{\sqrt{N}} \ket{\phi_+} + \ket{\phi_{k,-}} \right)  \left( \frac{1}{\sqrt{N}} \bra{\phi_+} + \bra{\phi_{k,-}}\right)  \\ 
        & = \frac{1}{N} \left( \ket{\phi_+}\bra{\phi_+} + \sum_{k=1}^{N} \ket{\phi_{k,-}}\bra{\phi_{k,-}} \right),
    \end{split}
\end{equation}
which implies
\begin{equation}
      P_U \left( \cdot \right) \geq \frac{1}{N} P_U \left( \text{error} \cap \text{herald} |  { \ket{\phi_+}\bra{\phi_+}}  \right).
\end{equation}
Therefore, the distilled error is bounded as
\begin{equation}
    \epsilon^\prime \geq \frac{P_U \left( \text{error} \cap \text{herald} |  { \ket{\phi_+}\bra{\phi_+}}  \right)}{p_0}\epsilon + \mathcal{O}\left( \epsilon^2 \right),
\end{equation}
or equivalently,
\begin{equation}
    \epsilon^\prime \geq \frac{P_U \left( \text{herald} | {\scriptstyle \ket{\phi_+}\bra{\phi_+} } \right) P_U \left( \text{error} | \text{herald}\cap {\scriptstyle \ket{\phi_+}\bra{\phi_+}}  \right)}{p_0}\epsilon.
\end{equation}
As shown in Reference\  \cite{saied2025general}, the state vector $\ket{\phi_+}$ behaves equivalently to a fully indistinguishable state under interference, which implies \\$P_U \left( \text{herald} | {\scriptstyle \ket{\phi_+}\bra{\phi_+} } \right) = p_0$, or 
\begin{equation}
    \epsilon^\prime \geq P_U \left( \text{error} | \text{herald}\cap { \ket{\phi_+}\bra{\phi_+}}  \right) \epsilon + \mathcal{O}\left( \epsilon^2 \right).
\end{equation}
The pseudo-indistinguishability of $\ket{\phi_+}$  ensures that all input modes are equally likely to contribute to the distilled photon, so that
\begin{equation}
    P_U \left( \text{error} | \text{herald}\cap { \ket{\phi_+}\bra{\phi_+}}  \right) = \frac{1}{N},
\end{equation}
thereby completing the proof.
\end{proof}

\section{Experimental methods}\label{App:ExpMethods}
\begin{figure}[h!]
    \centering
     \includegraphics[width=\columnwidth,  keepaspectratio]{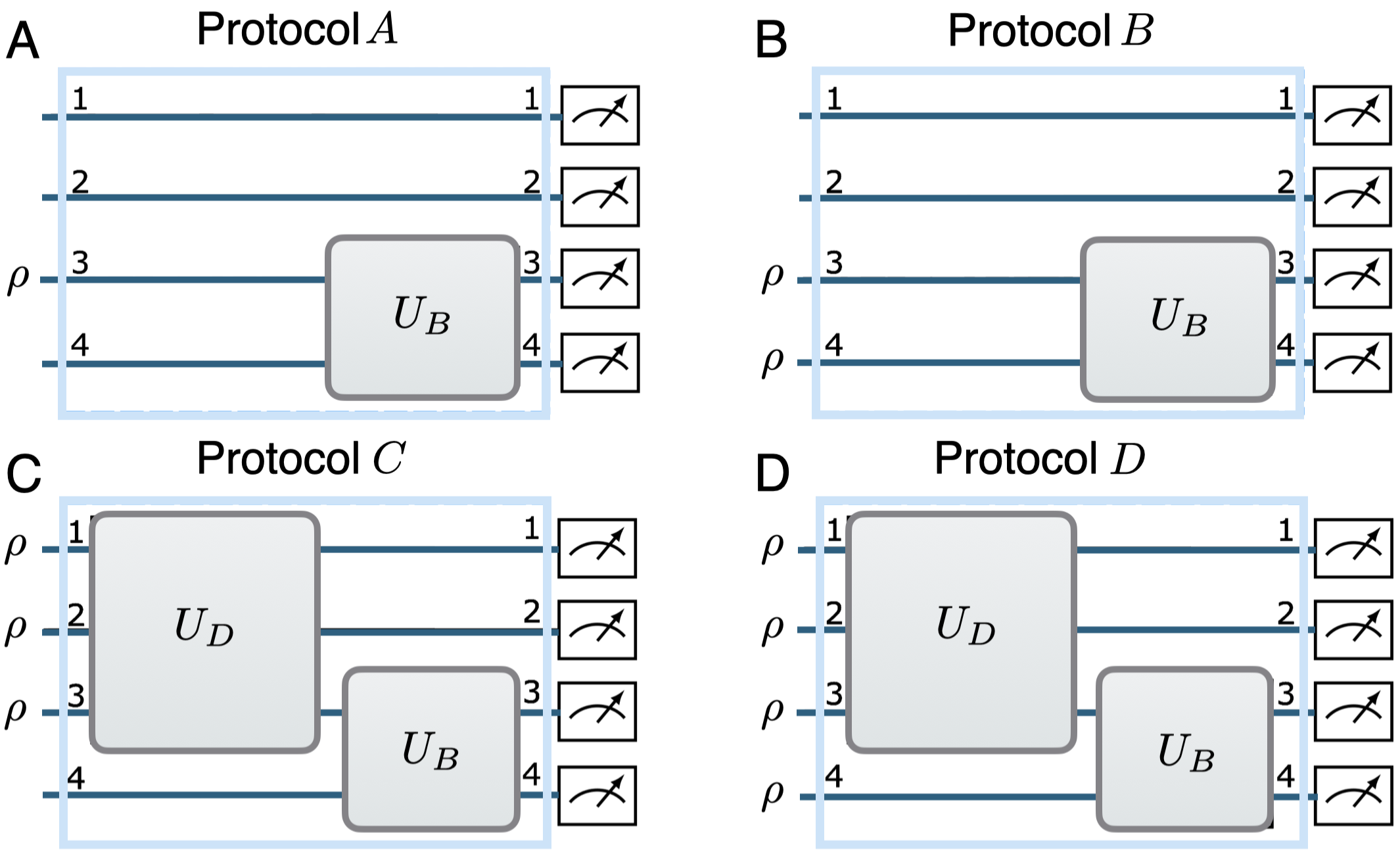}
    \caption{\textbf{Overview of the experimental (sub)protocols.} The $4\times 4$ block, highlighted by the blue dashed line, 
    is implemented in a four-mode partition on the quantum photonic processor. $\rho$ = injected photon; $U_D$ = distillation protocol \cite{marshall2022distillation}; $U_B$ = two-mode balanced beam splitter for HOM interference experiment.}
    \label{fig:expprotocols}
\end{figure}

We describe the experimental methods used to determine the indistinguishability and multiphoton errors before and after photon distillation. First, we introduce a series of measurement protocols (Figure\  \ref{fig:expprotocols}) used to obtain HOM and HBT correlators at zero time delay \cite{fischer2016dynamical,trivedi2020generation}, inspired by the measurement protocol introduced in Reference\  \cite{somhorst2025photon}. Second, we present the framework for relating these measured correlators $g_A$, $g_B$, $g_C$, and $g_D$ to the total and indistinguishability errors, as well as to multiphoton errors.  

First, we use Protocols $A$ and $B$ to characterise the photon properties \textit{prior} to distillation. Photons are injected into input mode 3 (and 4 for Protocol $B$) 
(Figure\  \ref{fig:expprotocols} a,b), and we integrate the single and coincidence counts over a measurement period $\Delta T$, using a coincidence window of $\Delta t = 2$ ns. We record the total single counts $N_3$ and $N_4$ in output modes 3 and 4, respectively, as well as the coincidence counts $N_{3,4}$ between outputs 3 and 4. In addition, we define $N_t$ as the number of trigger events, corresponding to the number of attempts to generate $N$ photons synchronously. We record this quantity using a separate channel that counts the excitation laser pulses, which is in postprocessing divided by $N$ to account for demultiplexing.\footnote{Due to data overflow, only one in $10^4$ trigger events is recorded. The reported counts were corrected for this undersampling in postprocessing.} Using the recorded single and coincidence counts, the correlators for Protocols $A$ and $B$ are calculated analogously \cite{grangier1986experimental} as
\begin{equation}
    g_{A (B)} = \frac{N_{3,4} N_t}{N_3N_4}. 
\end{equation}

Second, we use Protocols $C$ and $D$ to characterise the photon properties \textit{after} distillation. Photons are injected into input modes 1, 2 and 3 (as well as mode 4 for Protocol $D$) (Figure\   \ref{fig:expprotocols} c,d). Similarly to Protocols $A$ and $B$,  we integrate counts over a measurement period $\Delta T$ and use a coincidence window $\Delta t = 2$ ns. Successful photon-distillation events are postselected by requiring a coincidence detection in output modes 1 and 2, which heralds success \cite{marshall2022distillation}. Therefore, we record the coincidence counts $N_{1,2}$, $N_{1,2,3}$, $N_{1,2,4}$, and $N_{1,2,3,4}$. Using the recorded coincidence counts, the correlators for Protocols $C$ and $D$ are calculated through the following evaluation:
\begin{equation}
    g_{C (D)} = \frac{N_{1,2,3,4} N_{1,2}}{N_{1,2,3} N_{1,2,4}}. 
\end{equation}

Below, we describe how the measured $g_i$ values are used to infer the input and output errors of the imperfect single-photon state. We first note that the extraction of input and output multiphoton errors is relatively straightforward:
\begin{equation}
    \epsilon_{\text{multi}} = \frac{1}{2} g_A,
    \label{eq:g_in_def}
\end{equation}
and
\begin{equation}
    \epsilon_{\text{multi}}^\prime = \frac{1}{2} g_C.
    \label{eq:g_out_def}
\end{equation}

Extracting the indistinguishability errors requires multiple steps \cite{somhorst2025extracting}. We first compute the raw visibilities, accounting for an imperfect reflectivity $R_i$ in $U_{BS}$
\begin{equation}
    V_0 = \frac{R_1^2 + (1-R_1)^2 - g_B}{2R_1(1-R_1)},
    \label{eq:V0}
\end{equation}
and
\begin{equation}
    V_1 = \frac{R_2^2 + (1-R_2)^2 - g_D}{2R_2(1-R_2)}.
    \label{eq:V1}
\end{equation}

Second, we extract the input indistinguishability errors. Following Reference\  \cite{somhorst2025extracting}, we have \begin{equation}
	(1-\epsilon_{\text{tot}})^2 = V_0 + g_A,
\end{equation}
from which the total input error is 
\begin{equation}
	\epsilon_{\text{tot}} = 1 - \sqrt{V_0 + g_A}.
    \label{eq:eps_eff_in}
\end{equation}
Using Eq.\  (\ref{eq:eps_int_def}), the input indistinguishability error is then given by
\begin{equation}
	\epsilon_{\text{indist}} = 1 - \frac{\sqrt{V_0 + g_A}}{1 - \frac{1}{2}g_A}.
    \label{eq:eps_int_in}
\end{equation}

Analogously, we extract the output indistinguishability errors, which satisfy\footnote{Note that here we assume $\zeta = 1$, which does not significantly affect the interpretation of the results (see discussion around Eq. (\ref{eq:g_average_zeta})).}
\begin{equation}
    (1-\epsilon_{\text{tot}})(1-\epsilon_{\text{tot}}^\prime) = V_1 + \frac{1}{2}g_A + \frac{1}{2}g_C,
\end{equation}
from which the total output error is, 
\begin{equation}
	\epsilon_{\text{tot}}^\prime = 1 - \frac{V_1 + \frac{1}{2}g_A + \frac{1}{2}g_C}{\sqrt{V_0 + \frac{1}{2}g_A}},
    \label{eq:eps_eff_out}
\end{equation}
and the output indistinguishability error is given by
\begin{equation}
	\epsilon_{\text{indist}}^\prime = 1 - \frac{V_1 + \frac{1}{2}g_A + \frac{1}{2}g_C}{(1 - \frac{1}{2}g_C)\sqrt{V_0 + \frac{1}{2}g_A}}.
    \label{eq:eps_int_out}
\end{equation}

\section{Data processing and error propagation}\label{App:DataProcessing}
\begin{figure}[h!]
    \centering
     \includegraphics[width=\columnwidth,  keepaspectratio]{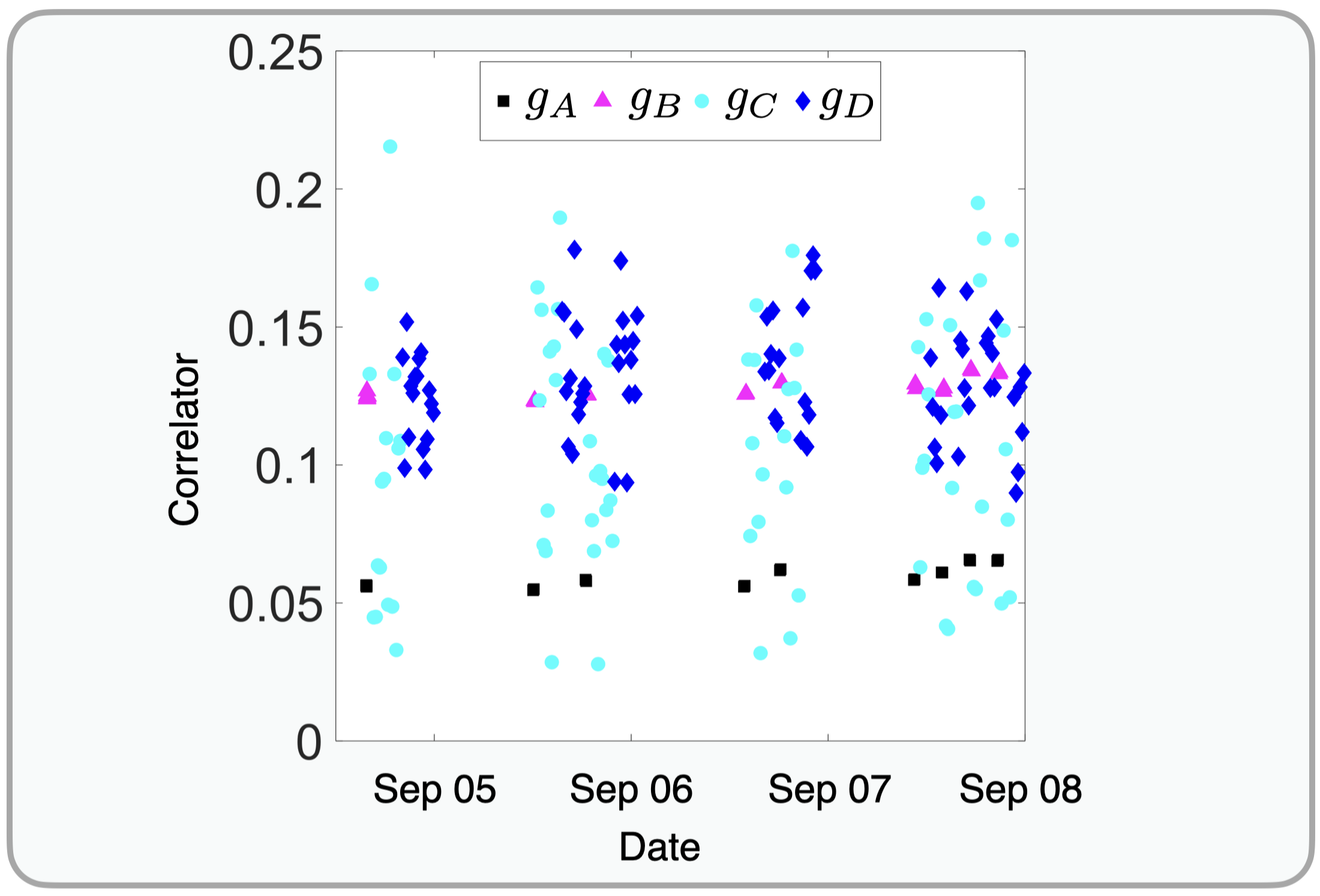}
    \caption{Timeline of correlator measurements.}
    \label{fig:TimeLineCorrelatorsMeasurements}
\end{figure}

We report on the statistical analysis of the raw data and the procedures used for error propagation. All measurements were collected between September 4, 2025, and September 7, 2025. During this period, we alternated between the different measurement protocols and repeated Protocols $C$ and $D$ particularly often, due to their low recorded multiphoton coincidence counts. The measurement integration time was $\Delta T = 300$ s for Protocols $A$ and $B$,\footnote{Except on September 4, where $\Delta T = 60$ s.} and $\Delta T = 900$ s for Protocols $C$ and $D$, and the corresponding sample correlators are presented in Figure\  \ref{fig:TimeLineCorrelatorsMeasurements}. Table \ref{tab:correlators} summarises the statistical analysis of the measured correlators, including the sample \emph{standard deviation} (SD) and \emph{standard error} (SE). Our subsequent analysis focusses on SE propagation to construct confidence intervals around the estimated photon errors.

\begin{table}[h]
\centering
\caption{Statistical summary of correlator measurements. SD = standard deviation, SE = standard error.} \label{tab:correlators}
\begin{tabular}{c c c c c}
\toprule
Correlator & Samples & Sample mean & Sample SD & Sample SE \\ 
\midrule
$g_{A}$ & $21$ & $0.0592$ & $0.0038$ & $0.0008$ \\
$g_{B}$ & $21$ & $0.1276$ & $0.0036$ & $0.0008$ \\
$g_{C}$ & $80$ & $0.104$ & $0.045$ & $0.005$ \\
$g_{D}$ & $80$ & $0.131$ & $0.021$ & $0.002$ \\
\botrule
\end{tabular}
\end{table}

We study the propagation of correlator uncertainty in the photon errors using standard error propagation through partial derivatives:
\begin{equation}
	\text{SE}_{f}^2 = {\sum_{i=1} \left( \frac{\partial f}{\partial x_i} \right)^2 (\text{SE}_{x_i})^2 }.
\end{equation}
We first observe that the correlator errors of Protocols $A$ and $C$ propagate in a straightforward manner to the error in the reported input and output multiphoton errors. We then analyse the intermediate visibilities, finding that\footnote{The error in the characterised reflectivity $R_i$ is neglected. Under a Poissonian noise model for the measured photon counts, Monte Carlo error propagation shows that the relative error in $R_1$ is less than 0.1\%, implying that correlator errors are the dominant contribution to the visibility uncertainties.}
\begin{equation}
	(\text{SE}_0)^2 = \left( \frac{1}{2R_1(1-R_1)} \right)^2 (\text{SE}_B)^2,
\end{equation}
and
\begin{equation}
	(\text{SE}_1)^2 = \left( \frac{1}{2R_2(1-R_2)} \right)^2 (\text{SE}_D)^2.
\end{equation}

Second, for the standard error in the total input error,
\begin{equation}
   (\text{SE}_{\epsilon_{\text{tot}}})^2 = \frac{(\text{SE}_0)^2 + (\text{SE}_A)^2}{V_0 + g_A},
\end{equation}
and for the input indistinguishability error
\begin{equation}
\begin{split}
    (\text{SE}_{\epsilon_{\text{indist}}})^2
     = \left( \frac{1}{(g_A-2)\sqrt{V_0 + g_A}} \right)^2 (\text{SE}_0)^2  & \\
    + \left( \frac{-2 V_0 - g_A - 2}{(g_A-2)^2 \sqrt{V_0 + g_A}} \right)^2 (\text{SE}_A)^2.
\end{split} 
\end{equation}

Similarly, for the standard error in the total output error,
\begin{equation}
\begin{split}
    (\text{SE}_{\epsilon_{\text{tot}}^\prime})^2
    = \left( \frac{2 V_1 + g_A + g_C}{4 (g_A + V_0)\sqrt{g_A + V_0}} \right)^2 (\text{SE}_0)^2  & \\
    + \left( \frac{- g_A -2 V_0 + g_C + 2 V_1}{4 (g_A + V_0)\sqrt{g_A + V_0}} \right)^2 (\text{SE}_A)^2 & \\
    + \left( \frac{-1}{\sqrt{g_A + V_0}} \right)^2 (\text{SE}_1)^2 & \\
    + \left( \frac{-1}{2 \sqrt{g_A + V_0}} \right)^2 (\text{SE}_C)^2,
\end{split} 
\end{equation}
and for the output indistinguishability error,
\begin{equation}
\begin{split}
    (\text{SE}_{\epsilon_{\text{indist}}^\prime})^2
    = \left( \frac{-2 V_1 - g_A - g_C}{2(g_C -2) (g_A + V_0)\sqrt{g_A + V_0}} \right)^2 (\text{SE}_0)^2  & \\
    + \left( \frac{g_A +2 V_0 - g_C - 2 V_1}{2(g_C -2) (g_A + V_0)\sqrt{g_A + V_0}} \right)^2 (\text{SE}_A)^2 & \\
    + \left( \frac{2}{(g_C - 2)\sqrt{g_A + V_0}} \right)^2 (\text{SE}_1)^2 & \\
    + \left( \frac{- g_A -2 V_1 -2}{(2-g_C)^2 \sqrt{g_A + V_0}} \right)^2 (\text{SE}_C)^2. 
\end{split} 
\end{equation}

Table \ref{tab:ResultsSummary} reports the propagated standard errors together with the corresponding 95\% confidence intervals.

\begin{table}[h]
\centering
\caption{Summary of standard errors and 95\% confidence intervals for the extracted parameters.} \label{tab:ResultsSummary}
\begin{tabular}{c c c c}
\toprule
Parameter & Value & SE & 95\% Confidence interval \\ 
\midrule
$V_0$ & $0.745$ & $0.002$ & $0.745 \pm 0.003$ \\
$V_1$ & $0.739$ & $0.005$ & $0.739 \pm 0.009$ \\
$\epsilon_{\text{multi}}$ & $0.030$ & $0.001$ & $0.030 \pm 0.001$ \\
$\epsilon_{\text{multi}}^\prime$ & $0.052$ & $0.003$ & $0.052\pm 0.005$ \\
$\epsilon_{\text{tot}}$ & $0.103$ & $0.002$ & $0.103 \pm 0.004$ \\
$\epsilon_{\text{tot}}^\prime$ & $0.084$ & $0.006$ & $0.084 \pm 0.012$ \\
$\epsilon_{\text{indist}}$ & $0.076$ & $0.001$ & $0.076 \pm 0.002$ \\
$\epsilon_{\text{indist}}^\prime$ & $0.034$ & $0.008$ & $0.034 \pm 0.015$ \\
\botrule
\end{tabular}
\end{table}

\section{Transfer matrix characterisation}\label{app:matrixCharacter}
We assess the accuracy of our implemented linear-optical transformations by characterising the experimentally realised transformation matrices. This characterisation serves two main purposes. First, it enables accurate extraction of the beam-splitter reflectivity used in the Hong-Ou-Mandel interference experiments, as deviations from ideal 50/50 splitting lead to systematic errors in visibility estimation and, consequently, in photon indistinguishability \cite{ollivier2021hong}. Second, it quantifies the fidelity of the implemented distillation matrix. This information is particularly important because 
the photon-distillation circuit under study is not fault-tolerant, although it exhibits some robustness to circuit imperfections \cite{marshall2022distillation}. Therefore, characterisation of the experimental distillation circuit is essential to implement strategies for matrix accuracy improvement.

The matrix amplitudes of an implemented linear-optical transformation $T$ are characterised using only single-photon measurements \cite{laing2012super}. In this procedure, photons are injected into input port $i$, while photon detection counts are recorded simultaneously at all output ports $j$. The resulting normalised count distribution corresponds to the $i^{\text{th}}$ row of the transformation $T$, with the counts proportional to $|T_{i,j}|^2$. Each single-photon measurement is integrated over a duration of 5 s, corresponding to a normalisation factor of $S_{\text{norm}} = 3.955 \times 10^8$, which represents the number of single-photon source excitations.  The recorded photon counts $S_{i,j}$ for the distillation experiment are as follows:
\begin{equation}
    S = 
    \begin{bmatrix}
    2688386 & 2792549 & 1340100 & 1052599 \\
    1896852 & 2623508 & 1109520 & 904938 \\
    3110744 & 3153085 & 1989633 & 1570463 \\
    17966 & 15195 & 4762700 & 4338881 
    \end{bmatrix}. 
    \label{eq:S_recorded}
\end{equation}

The characterised transformation $T$ is generally non-unitary due to accumulated losses from photon generation to detection. Assuming uniform on-chip losses,\footnote{This assumption follows from the rectangular geometry of the interferometric network, designed to yield balanced losses \cite{clements2016optimal}.} we model the total system as an implemented unitary transformation $U_{\text{exp}}$ sandwiched between two sub-unitary diagonal matrices accounting for input and output losses \cite{laing2012super}
as
\begin{equation}
	T = D_{\text{in}} | U_{\text{exp}} | D_{\text{out}}.
    \label{eq:T_decomposition}
\end{equation}
The unitary constraint on $U_{\text{exp}}$  yields a solvable set of nonlinear equations, allowing extraction of the matrix amplitudes of $U_{\text{exp}}$ (indicated by $|U_{\text{exp}}|$) and the corresponding loss matrices $D_{\text{in}} $ and $D_{\text{out}}$.\footnote{ $ D_{\text{in}} $ and $D_{\text{out}}$ are not unique due to the commutation of loss with unitary transformations \cite{oszmaniec2018classical}.} For the distillation experiment, we obtain the following result:
\begin{align}
\left[ \begin{smallmatrix}
0.0824 & 0.0840 & 0.0582 & 0.0516 \\
0.0693 & 0.0814 & 0.0530 & 0.0478 \\
0.0887 & 0.0893 & 0.0709 & 0.0630 \\
0.0067 & 0.0062 & 0.1097 & 0.1047
\end{smallmatrix} \right] \notag
&= 
\left[ \begin{smallmatrix}
0.3568 & 0 & 0 & 0 \\
0 & 0.3242 & 0 & 0 \\
0 & 0 & 0.3991 & 0 \\
0 & 0 & 0 & 0.3909
\end{smallmatrix} \right] \\ \notag
& \quad \cdot
\left[ \begin{smallmatrix}
0.5992 & 0.5731 & 0.4032 & 0.3872 \\
0.5540 & 0.6114 & 0.4038 & 0.3952\\
0.5762 & 0.5443 & 0.4392 & 0.4228 \\
0.0447 & 0.0386 & 0.6939 & 0.7177
\end{smallmatrix} \right] \notag \\
& \quad \cdot
\left[ \begin{smallmatrix}
0.3856 & 0 & 0 & 0 \\
0 & 0.4110 & 0 & 0 \\
0 & 0 & 0.4046 & 0 \\
0 & 0 & 0 & 0.3734
\end{smallmatrix} \right].
\label{eq:MatrixCharacterFull}
\end{align}

The distillation experiment is implemented as a three-mode distillation matrix, with the third output mode concatenated to a two-mode beam splitter to enable interference between the distilled and non-distilled photons (see Reference\  \cite{somhorst2025photon}, Appendix B, Figure\ 7). We fit the inferred unitary matrix amplitudes $|U_{\text{exp}}|$ using this model, yielding $U_{\text{model}}$. First, a beam splitter with variable reflectivity $R_2$ is used to minimise the off-diagonal elements in the fourth row and column of $|U_{\text{exp}}|$.\footnote{Due to leakage from input 4 to outputs 1 and 2, this off-diagonals are only approximate zero, and the resulting $3\times 3$ submatrix is renormalised.} Second, it is informally known that the matrix amplitudes of a three-mode unitary matrix fix the phases (up to an overall sign ambiguity \cite{laing2012super}), eliminating the need for two-photon interference measurements. We find that
\begin{equation}
    |U_{\text{model}}| = 
    \begin{bmatrix}
    0.5998 & 0.5735 & 0.4012 & 0.3879 \\
    0.5546 & 0.6118 & 0.4055 & 0.3921 \\
    0.5768 & 0.5448 & 0.4376 & 0.4231 \\
    0 & 0 & 0.6951 & 0.7189
    \end{bmatrix},
\end{equation}
resulting in a linear-optical circuit (amplitude) fidelity of \\$\mathcal{F}_{\text{fit}} = \frac{1}{4}{\rm tr} ( |U_{\text{exp}}^\dagger| |U_{\text{model}}| ) = 0.9996$, validating the concatenated distillation experiment model. For the distillation experiment, we find a beam-splitter reflectivity of $R_2~=~0.517$.   

\begin{figure}[h!]
    \centering
     \includegraphics[width=\columnwidth]{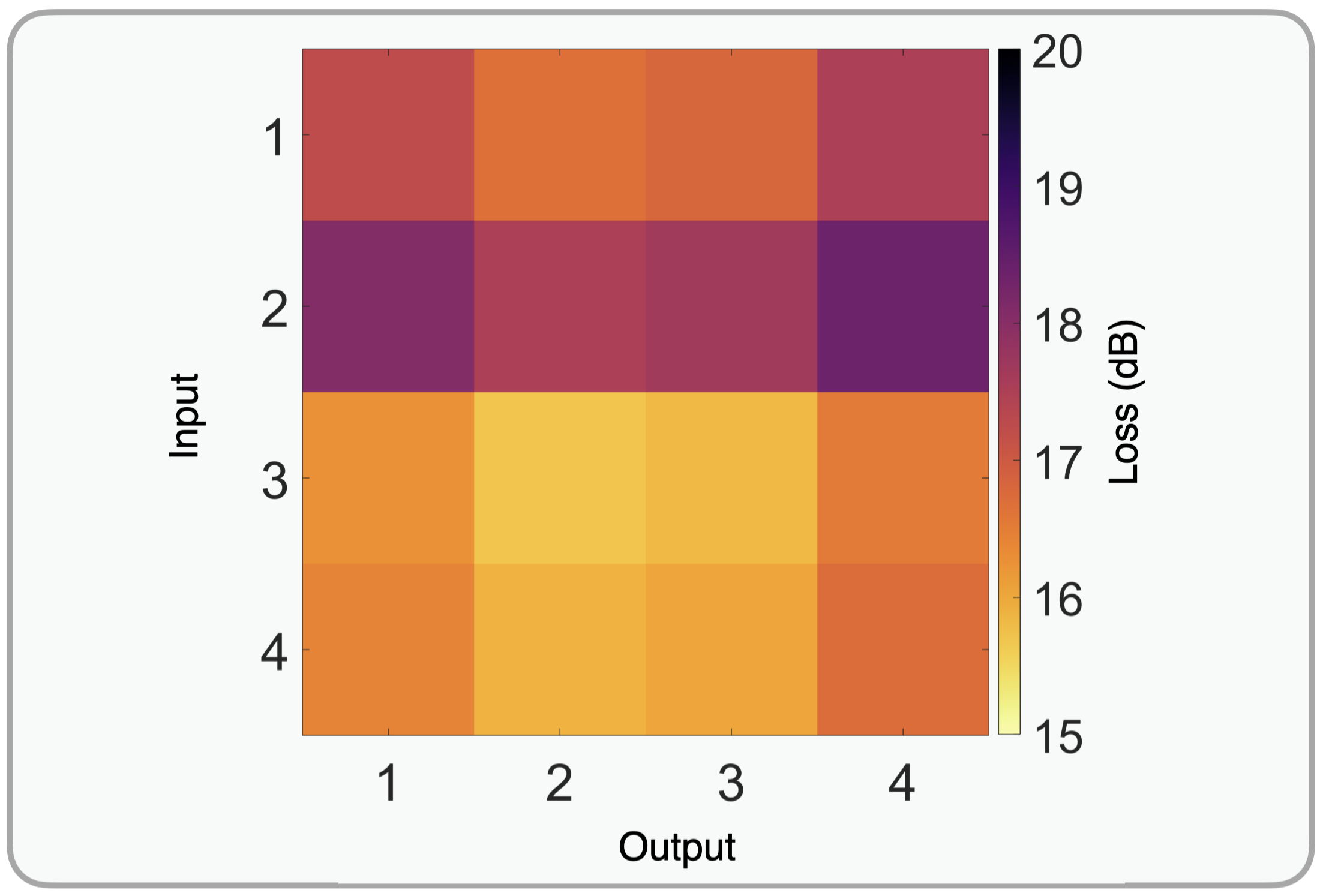}
    \caption{\textbf{End-to-end single-photon transmission loss for each input–output pair.}}
    \label{fig:system_loss_deduced}
\end{figure}

The reconstructed distillation matrix is,
\begin{equation}
    U_D^{\text{exp}} = 
    \left[ \begin{smallmatrix}
        0.5998+0.0000i & 0.5735 + 0.0000i & 0.5580 + 0.0000i \\
        0.5546 + 0.0000i & -0.3530 - 0.4997i & -0.2333 +0.5135i \\
        0.5768 +0.0000i & -0.2569 + 0.4804i & -0.3560 - 0.4937i
    \end{smallmatrix} \right].
    \label{eq:MatrixU_D_Exp}
\end{equation}
Ideally, the distillation matrix is given by,
\begin{equation}
    U_D^{\text{th.}} = 
    \left[ \begin{smallmatrix}
        0.5774 + 0.0000i & 0.5774 + 0.0000i & 0.57774 + 0.0000i \\
        0.5774 + 0.0000i & -0.2887 - 0.5000i & -0.2887 + 0.5000i \\
        0.5774 + 0.0000i & -0.2887 + 0.5000i & -0.2887 - 0.5000i
    \end{smallmatrix} \right],
\end{equation}
which corresponds to a distillation circuit (full) fidelity of 
\begin{equation}
    \mathcal{F}_D = \frac{1}{3} {\rm tr} \left( \left (U_D^{\text{th}} \right )^\dagger U_D^{\text{exp}} \right) = 0.9982. \label{eq:fidelity}
\end{equation}
Furthermore, we estimate the single-photon end-to-end transmission efficiency $\eta_{i,j}$. We further note that 
\begin{equation}
\eta_{i,j} = \left ( D_{\text{in},i,i} \right )^2 \left ( D_{\text{out},j,j} \right) ^2. 
\end{equation}
We plot the corresponding end-to-end loss, $-10 \log{ \eta_{i,j} }$, for the processor partition used in 
Figure\ \ref{fig:system_loss_deduced}. From this figure, we find an average transmission efficiency of $\eta = 0.021 \pm 0.004$.

A similar characterisation procedure is applied to the reference experiment to determine the reflectivity $R_1$. Since only ports 3 and 4 are used in the reference Protocols $A$ and $B$, we report only the corresponding input–output counts for these ports.  The recorded photon counts $S^ {\text{ref}}_{i,j}$ for the reference experiment are presented in Eq.\  (\ref{eq:S_recorded_ref}): 
\begin{equation}
    S^ {\text{ref}} = 
    \begin{bmatrix}
     \times &  \times &  \times&  \times \\
     \times &  \times &  \times&   \times\\
     \times &  \times & 4872300 & 4228197 \\
     \times &  \times & 4958917 & 4197710 \\ 
    \end{bmatrix}. 
    \label{eq:S_recorded_ref}
\end{equation}
Applying the decomposition of Eq.\ (\ref{eq:T_decomposition}) to the reference experiment yields (showing only the relevant $2\times 2$  submatrix)
\begin{align}
\left[ \begin{smallmatrix}
0.1110 & 0.1034 \\
0.1120 & 0.1030 
\end{smallmatrix} \right] \notag
&= 
\left[ \begin{smallmatrix}
0.3809 & 0 \\
0 & 0.3819
\end{smallmatrix} \right] \\ \notag
& \quad \cdot
\left[ \begin{smallmatrix}
0.7049 & 0.7093 \\
0.7093 & 0.7049
\end{smallmatrix} \right] \notag \\
& \quad \cdot
\left[ \begin{smallmatrix}
0.4134 & 0 \\
0 & 0.3827
\end{smallmatrix} \right].
\label{eq:MatrixCharacterReference}
\end{align}
From this decomposition, we obtain $\sqrt{R_1}~=~0.7049$, which implies $R_1~=~0.497$.

\section{Resource estimation for logical qubits in single-photon units}\label{app:PhotonResourceEstimate} 
We estimate the single-photon resource cost for encoding one logical qubit using the framework of References \cite{saied2025general,somhorst2025photon}. The physical noise model includes only Pauli errors generated by fusion measurements between photons with indistinguishability error $\epsilon_{\text{indist}}$, resulting in \cite{rohde2006error, somhorst2025photon}
\begin{equation}
	p_{\text{error}} = \frac{1}{2} \epsilon_{\text{indist}} + \mathcal{O}\left( \epsilon_{\text{indist}}^2 \right).
\end{equation}
We consider a measurement-based quantum computing architecture in which a surface code of distance $d$ encodes one logical qubit using $d^3$ resource states \cite{bombin2023logical}. The logical error rate is given just so \cite{fowler2012surface}: 
\begin{equation}
	p_{\text{L}} = \left( \frac{p_{\text{error}}}{p_{\text{th}}} \right)^{d/2},
\end{equation}
where $p_{\text{th}}$ denotes the Pauli-error threshold. Each resource state incurs a fixed photon cost $b$, for a total photon cost of 
\begin{equation}
	C_{\text{L}} = b d^3.
\end{equation}	
A hybrid combination of QEC and photon distillation amplifies the tolerable indistinguishability error and reduces resource requirements below threshold \cite{somhorst2025photon}. A photon-distillation scheme of size $N$ suppresses the Pauli error rate $p_{\text{error}}$ by a factor of $N$, while increasing the logical-qubit overhead $C_{\text{L}}$ by a factor of $4N$ \cite{saied2025general,somhorst2025photon}. This scaling is valid in the asymptotic regime, where at most one out of $N$ interfered photons is distinguishable. To evaluate this assumption, we consider the relative probability of multiple versus single distinguishable photons \cite{somhorst2025photon}:
\begin{equation}
    \frac{P(>\text{1 error})}{P(\text{1 error})} = \frac{1-(1-\epsilon_{\text{indist}})^N -N\epsilon_{\text{indist}}(1-\epsilon_{\text{indist}})^{N-1}}{N\epsilon_{\text{indist}}(1-\epsilon_{\text{indist}})^{N-1}}.
\end{equation}
For the indistinguishability errors discussed in the main text, \\${P(>\text{1 error})}/{P(\text{1 error})}~\leq~{1}/{50}$, justifying the photon-distillation scaling assumptions.

We analyse a basic QEC code that consumes six-ring resource states, with a Pauli-error threshold of $p_{\text{th}}~=~2.1~\times~10^{-3}$ in the absence of loss \cite{bombin2023increasing,saied2025general}. Following Reference\  \cite{bartolucci2025comparison}, a six-ring resource state is assembled from 20 three-qubit GHZ (3-GHZ) states using type-I fusions. This cost follows from the different fusion steps: two initial 3-GHZ states produce a four-qubit cluster, with a cost of $(1 + 1)\times2^1 = 4$ 3-GHZs, and two further fusion attempts with a single 3-GHZ state results in the six-ring resource state, yielding a cost of $(4 + 1)\times 2^2 = 20$ 3-GHZs. Each 3-GHZ state is generated using an average of $6 \times 54 = 324$ photons under the probabilistic scheme of Refs.\  \cite{gubarev2020improved,chen2024heralded}, giving $b = 6480$ photons. Setting the target logical error to $p_{\text{L}} = 10^{-10}$ \cite{GoogleUnderThresholdCodes}, the required code distance $d$ (assumed continuous) is inferred, allowing calculation of the photon resource cost for one logical qubit as presented in the main text. Estimated indistinguishability errors of state-of-the-art single-photon sources are summarised in Table \ref{tab:eps_SPS} for deterministic and probabilistic single-photon sources. Errors refer to the mismatch between photons emitted by different devices.

\begin{table}[h]
\centering
\caption{Estimated state-of-the-art mutual indistinguishability errors for photons emitted by different probabilistic (P) and deterministic (D) single-photon sources.} \label{tab:eps_SPS}
\begin{tabular}{l l l l l}
\toprule
Point & Year & Authors & Type & $\epsilon_{\text{indist}}$ \\ 
\midrule
A & 2025 & PsiQuantum team \cite{PsiQuantumPlatform} & P & $2.5 \times 10^{-3}$ \\
B & 2020 & Paesani et al. \cite{paesani2020near} & P & $6.5 \times 10^{-3}$ \\
C & 2022 & Zhai et al. \cite{zhai2022quantum} & D & $3.6 \times 10^{-2}$ \\
\botrule
\end{tabular}
\end{table}

\section{Overview of resource scaling across different photon-distillation schemes}\label{app:ScalingOverview}
We quantify the asymptotic resource cost $C$ required to distil a photon with indistinguishability error $\epsilon_{\text{indist}}^\prime < \epsilon_{\text{indist}}$. This scaling is characterised by the $\gamma$-exponent, via 
\begin{equation}
	C = \mathcal{O} \left( \left( \frac{\epsilon_{\text{indist}}}{\epsilon_{\text{indist}}^\prime} \right)^\gamma \right).
\end{equation}

The first proposed family of photon-distillation schemes by Sparrow et al. \cite{sparrow2018quantum} exploit enhanced bunching of indistinguishable bosons within a single optical mode in combination with probabilistic photon-number subtraction. However, the probability of obtaining single-mode bunched outcomes falls rapidly with scheme size \cite{yung2019universal}, rendering such generalised schemes inefficient. Alternatively, concatenated two-photon schemes yield $\gamma = 3$.\footnote{Each iteration of the two-photon scheme asymptotically requires eight photons to halve the error. To suppress the error by a factor $\frac{\epsilon_{\text{indist}}}{\epsilon_{\text{indist}}^\prime} = 2^r$, $r$ iterations are needed, consuming a total of $8^r = (2^r)^3 = \left( \frac{\epsilon_{\text{indist}}}{\epsilon_{\text{indist}}^\prime} \right) ^3$ photons.}

Marshall et al. \cite{marshall2022distillation} have reduced this to $\gamma = 2$, by harnessing enhancements in multimode statistics rather than single-mode statistics to
devise a concatenated three-photon scheme. 
More recently, Somhorst et al. \cite{somhorst2025photon} and Saied et al. \cite{saied2025general} 
have independently identified that these multimode enhancements arise from structured interferometers obeying zero-transmission laws \cite{tichy2010zero}, the multiphoton analogue of the Hong–Ou–Mandel interference \cite{hong1987measurement}. This structure enabled generalisation to single-step schemes achieving \textit{linear} scaling ($\gamma = 1$).

To our knowledge, these works encompass all photon-distillation schemes reported to date. Table \ref{tab:DistillationSchemesSummary} summarises their key characteristics.

\begin{table}[h]
\centering
\caption{Summary of asymptotic resource scaling for photon-distillation schemes. The table summarises key improvements in overhead exponent $\gamma$-values for the distillation of indistinguishable photons.} \label{tab:DistillationSchemesSummary}
\begin{tabular}{l l c}
\toprule
Year & Authors & Overhead \\
exponent $\gamma$ \\ 
\midrule
2017 & Sparrow et al. \cite{sparrow2018quantum} & $ 3$ \\
2022 & Marshall et al. \cite{marshall2022distillation} & $2$ \\
2025 & Somhorst et al. \cite{somhorst2025photon} & $ 1$ \\
2025 & Saied et al. \cite{saied2025general} & $ 1$ \\
\botrule
\end{tabular}
\end{table}

\end{appendices}

\bibliography{ReferencesArXiv}

\end{document}